\documentclass{IEEEtran}
\IEEEoverridecommandlockouts
\usepackage{cite}
\usepackage{amsthm}
\usepackage{amsmath,amssymb,amsfonts}
\usepackage{algorithm}
\usepackage{algorithmicx}
\usepackage{algpseudocode}
\usepackage{graphicx}
\usepackage{textcomp}
\usepackage{xcolor}
\usepackage{bm}
\usepackage{graphicx,epstopdf,psfrag,url,cite,xcolor,multirow,float}
\usepackage[font=footnotesize]{subfig}

\graphicspath{{figures/}}
\def\BibTeX{{\rm B\kern-.05em{\sc i\kern-.025em b}\kern-.08em
    T\kern-.1667em\lower.7ex\hbox{E}\kern-.125emX}}
\begin{document}
\title{DRL-driven Online Optimization for Joint Traffic Reshaping and Channel Reconfiguration in RIS-assisted Semantic NOMA Communications}
\author{Songhan Zhao, Shimin Gong,~\IEEEmembership{Member,~IEEE}, Bo Gu,~\IEEEmembership{Member,~IEEE}, Zehui Xiong,~\IEEEmembership{Senior Member,~IEEE}, \\Ping Wang,~\IEEEmembership{Fellow,~IEEE}, and Kaibin Huang,~\IEEEmembership{Fellow,~IEEE}

\thanks{Songhan Zhao, Shimin Gong, and Bo Gu  are with the School of Intelligent Systems Engineering, Sun Yat-sen University, China (e-mail: zhaosh55@mail2.sysu.edu.cn; gongshm5@mail.sysu.edu.cn; gubo@mail.sysu.edu.cn). Zehui Xiong is with the School of Electronics, Electrical Engineering and Computer Science (EEECS), Queen's University Belfast, United Kingdom (e-mail: z.xiong@qub.ac.uk). Ping Wang is with the Department of Electrical Engineering and Computer Science, Lassonde School of Engineering, York University, Canada (e-mail: pingw@yorku.ca). Kaibin Huang is with the Department of Electrical and Electronic Engineering, The University of Hong Kong, Hong Kong (e-mail: huangkb@eee.hku.hk).
}
}
\maketitle
\thispagestyle{empty}

\begin{abstract}
This paper explores a reconfigurable intelligent surface (RIS)-assisted and semantic-aware wireless network, where multiple semantic users (SUs) transmit semantic information to an access point (AP) using the non-orthogonal multiple access (NOMA) method. The RIS reconfigures channel conditions, while semantic extraction reshapes traffic demands, providing enhanced control flexibility for NOMA transmissions. To enable efficient long-term resource allocation, we propose a deferrable semantic extraction scheme that can distribute the semantic extraction tasks across multiple time slots. We formulate a long-term energy efficiency maximization problem by jointly optimizing the RIS's passive beamforming, the SUs' semantic extraction, and the NOMA decoding order. Note that this problem involves multiple and coupled  control variables, which can incur significant computational overhead in time-varying network environments. To support low-complexity online optimization, a deep reinforcement learning (DRL)-driven online optimization framework is developed. Specifically, the DRL module facilitates the adaptive selection and optimization of the most suitable option from traffic reshaping, channel reconfiguration, or NOMA decoding order assignment based on the dynamic network status. Numerical results demonstrate that the deferrable semantic extraction scheme significantly improves the long-term energy efficiency. Meanwhile, the DRL-driven online optimization framework effectively reduces the running time while maintaining superior learning performance compared to state-of-the-art methods.
\end{abstract}
\begin{IEEEkeywords}
Semantic communication, RIS, NOMA, traffic shaping, DRL.
\end{IEEEkeywords}
\section{Introduction}
The massive interconnection of Internet of Things (IoT) devices generates an enormous surge in wireless traffic demand, necessitating more efficient technologies to meet increasing transmission requirements~\cite{Shi-pro2024}. The conventional multiple access methods, such as time division multiple access (TDMA) and frequency division multiple access (FDMA), allocate orthogonal wireless resources to IoT devices to avoid co-channel interference.  However, as the network scale expands, these orthogonal multiple access methods become increasingly inefficient, as the finite spectrum resources cannot be effectively allocated among a large number of wireless devices~\cite{Ahmed-2024icst}. To address this challenge, non-orthogonal multiple access (NOMA) has emerged as a key solution to multiple access.
The NOMA method allows multiple users to simultaneously share the same spectrum resources by leveraging power-domain multiplexing~\cite{Ding-2022pro}. The receiver employs the successive interference cancellation (SIC) method to iteratively decode each user's signal from the superimposed signals. However, the effectiveness of SIC heavily depends on maintaining a sufficient power difference between the NOMA users~\cite{Pei-2022jsac}. This motivates employing the channel reconfiguration techniques to dynamically manipulate signal power levels for multiple users and thus improve their NOMA transmission efficiency.

The emerging reconfigurable intelligent surface (RIS) has been demonstrated as an efficient technique for reconfiguring wireless channels. The RIS consists of a large array of passive reflecting elements that can induce phase shifts on the incident signals~\cite{An-iwc2024}. By designing the passive beamforming, the RIS can dynamically reconfigure the channel conditions on demand~\cite{Gong-2020icst}. Motivated by this, the integration of RIS and NOMA is expected to provide significant advantages in overcoming the limitations of conventional wireless networks. The RIS can be deployed around NOMA users to better adjust their signals into different power domains, which efficiently improves the design flexibility for the SIC decoding~\cite{Hou-2020jsac}. Besides the terrestrial RIS, RIS can be also mounted on unmanned aerial vehicles (UAVs) forming the aerial-RIS (ARIS) systems. By exploiting the UAV's mobility, the ARIS can provide dynamic 3D beamforming and expand service coverage for ground users~\cite{Su-2022TCOM}. However, as the network traffic surges and the wireless environments grow increasingly complex, solely relying on RIS to control channel conditions may not be sufficient to meet the dynamic traffic demands of NOMA users. This highlights the need for traffic-aware technologies capable of dynamically adapting to real-time traffic variations in the NOMA transmissions.

The semantic communication, as a novel paradigm, has potential to surpass the transmission capacity limits in conventional wireless networks. By extracting the underlying meanings or goals from raw content, the semantic communication can reduce the need for transmitting redundant symbols~\cite{Yang-2023icst}. Compared to the Shannon paradigm that transmits the complete symbols of raw content, the semantic communication can transmit only important information and thus significantly improve the spectral efficiency. Building on this, integrating the semantic technique into NOMA transmissions can effectively address the ``early-late" issue, where earlier-decoded users experience stronger interference than those decoded later~\cite{Ding-2020cl}. By adaptively adjusting the semantic extraction depth, it enables the on-demand reshaping of the NOMA users' traffic demands to match with their transmission capacities. While the semantic extraction can reduce the transmission overhead, it also imposes the additional processing complexity during the extraction and recovery stages at the semantic users (SUs) and the access point (AP), respectively. As such, it is important to balance the trade-off between the semantic extraction efficiency and the transmission efficiency.
\subsection{Motivations and Challenges}
Inspired by the advantages of RIS and semantic communication, we aim to integrate both techniques to improve the NOMA transmission efficiency in scenarios with dynamic traffic and channel variation. In conventional NOMA wireless networks, the transmission performance is directly influenced by the decoding order of NOMA users. Typically, the NOMA decoding order is determined by sorting NOMA users' channel gains to satisfy the SIC requirements. However, this channel-based ordering struggles to meet the time-varying traffic demands in dynamic wireless environments. Therefore, it becomes essential to consider both channel conditions and traffic demands to design an adaptive NOMA decoding order. To this end, we explore an RIS-assisted semantic-aware NOMA transmission, where the RIS reconfigures channel conditions while the semantic extraction flexibly reshapes the users' traffic demands. By dynamically adapting the RIS's passive beamforming and semantic extraction depth, the NOMA users are expected to adjust their decoding order more flexibly and improve the transmission efficiency significantly. Although semantic extraction reduces transmission load, it incurs additional energy overhead. Thus, this paper aims to enhance the overall energy efficiency of the system. 

Another challenge is to design lightweight optimization algorithms for  online execution. Typically, the performance maximization problems in the RIS-assisted NOMA wireless networks can be solved by employing the model-based optimization methods. One common optimization approach is to decompose the complex original problem into multiple subproblems and then iteratively solve them with the approximate methods~\cite{Zhao-wcnc2025}, such as the semidefinite relaxation (SDR) and successive convex approximation (SCA). However, as the network scales, the complexity of these model-based optimization methods increases significantly, which makes them difficult to implement in practical scenarios. Moreover, the model-based optimization methods generally rely on perfect network information. However, due to the difficulty in obtaining the future information, these methods are effective at improving the short-term performance but struggle to optimize the long-term performance. Besides the model-based optimization, another common choice is the model-free deep reinforcement learning (DRL) methods. The DRL agents can adapt control variables by interacting with the environment without the need for the model information. The actor networks of the agents can quickly respond to the dynamic environment and make real-time decisions.  However, as the dimensionality of control variables increases, it becomes increasingly challenging for the DRL to learn all variables and achieve the optimal solutions with limited training samples in a time efficient manner.
\subsection{Solutions and Contributions}
In this paper, we investigate a novel RIS-assisted semantic-aware NOMA wireless network, where multiple SUs extract semantic information from raw data and then transmit it to the AP with the assistance of an RIS. The RIS reconfigures the channel conditions, while the semantic extraction reshapes the traffic demands of the SUs, jointly controlling the NOMA decoding order and improving the long-term transmission efficiency. We aim to maximize the long-term energy efficiency by jointly optimizing the RIS's passive beamforming, the semantic extraction, and the NOMA decoding order. To obtain initial insights from the proposed model, we first consider a special case that focuses on per-slot performance enhancement. We propose an alternating optimization (AO) method for the per-slot problem and design the model-based approximate methods for the decomposed subproblems. Building on this, we initially validate the effectiveness of the proposed joint traffic reshaping and channel reconfiguration (JTAC) scheme in the NOMA transmissions. However, the model-based optimization methods struggle to account for the long-term performance. Moreover, their high complexity characteristic also makes them challenging to execute online in dynamic scenarios.

To tackle the challenges above, we first develop a deferrable semantic extraction scheme, where semantic extraction tasks can be distributed across multiple time slots depending on dynamic network states. To address the long-term optimization problem, we design a DRL-driven online optimization framework. We consider that the optimization module does not need to obtain accurate solutions, as the DRL module can compensate for inaccuracies during the long-term interaction. As such, in each time slot, the system only needs to select the most suitable optimization option from traffic reshaping, channel reconfiguration, or NOMA decoding order assignment based on the current network status. This effectively avoids the alternating iterations in the optimization module, and thus reduces the complexity significantly. The main contributions of the paper are summarized as follows:

\begin{itemize}
\item \emph{RIS-assisted Semantic-aware NOMA Transmissions}:
We explore a novel RIS-assisted semantic-aware NOMA transmission scheme, which can effectively enhance the adaptability to scenarios with dynamic traffic and channel variation. The RIS reconfigures the channel conditions and the semantic extraction reshapes the traffic demands, providing greater control flexibility in the NOMA transmissions. The long-term energy efficiency is maximized by jointly optimizing the RIS's passive beamforming, the semantic extraction, and the NOMA decoding order.
\item \emph{Deferrable Semantic Extraction and Adaptive Access}:
To improve long-term energy efficiency, we propose a deferrable semantic extraction scheme that enables the SUs to intelligently distribute the semantic extraction tasks across multiple time slots and adaptively join the NOMA transmissions. This  effectively reduces the competition for the NOMA transmissions among multiple SUs, while the cross-slot semantic extraction better emphasizes the long-term performance enhancement.
\item \emph{DRL-driven Online Optimization Framework}: To reduce computational complexity in long-term optimization, we design a DRL-driven online optimization framework. Specifically, the optimization module focuses on the per-slot transmission efficiency, while the DRL method better improves the long-term performance. The DRL module enhances the optimization process by dynamically identifying and selecting the optimal option from traffic reshaping, channel reconfiguration, or NOMA decoding order assignment, thereby significantly reducing computational complexity. Simulations demonstrate that the DRL-driven online optimization framework efficiently improves long-term energy efficiency and reduces time overhead compared to benchmark methods.
\end{itemize}

The rest of the paper is organized as follows: We discuss the related work in Section~\ref{relate-work}. The system model is presented in Section~\ref{simultaneous-extraction-transmission} and a model-based optimization method is developed in Section~\ref{pure-opti-method} to validate the effectiveness of the JTAC scheme for NOMA transmissions. To improve the long-term performance, we further propose a deferrable semantic extraction scheme and a DRL-driven online optimization framework in Section~\ref{deferred-PGOO}. Finally, the proposed methods are evaluated in Section~\ref{results}, and conclusions are drawn in Section~\ref{conclusions}. {Key notations used in this paper are summarized in Table~\ref{Notation}.}
\begin{table}[t]
\caption{Summary of key notations.} \label{Notation}\normalsize
	\centering
	\begin{tabular}{|l|l|}
		\hline
        Symbol&Description\\
        \hline
        $t$ & Index of the time slot \\
        $k,k'$ & Indices of the SUs \\
        $p_k$ & Transmit power of SUs\\
        $\boldsymbol{\theta}$ & RIS's passive beamforming \\
        $\pi_{k,k'}$ & NOMA decoding order \\
        $\rho_k$ &  Semantic extraction depth \\
        $D_k$ & Data size for semantic extraction\\
        $Z_k$ & Data size for semantic transmission\\
        $\psi_k$ &  Transmission scheduling strategy\\
        $m$ &  Optimization selection strategy\\
        $\sigma^2$ & Background noise at the AP \\
        $S_k$ & Semantic capacity in Section~\ref{simultaneous-extraction-transmission}-\ref{pure-opti-method}\\
        $\widehat{S}_k$ & Semantic capacity in Section~\ref{deferred-PGOO} \\
        $S_o$ & Sum semantic capacity \\
        $B_{r,k}$ &  Data dynamic in Section~\ref{simultaneous-extraction-transmission}-\ref{pure-opti-method}\\
        $\widehat{B}_{r,k}$, $\widehat{B}_{s,k}$ & Data dynamics in Section~\ref{deferred-PGOO}\\
        $W_{e,k}$, $W_{r,k}\!\!\!\!$ & Computation load of semantic control \\
        $E_{s,k}$& Energy consumption of semantic control  \\
        $E_{t,k}$ & Energy consumption in transmission \\
        $E_o$ & Overall energy consumption \\
        $\eta$, $\widehat{\eta}$ & Energy efficiency in Section~\ref{pure-opti-method} and \ref{deferred-PGOO} \\
        \hline
	\end{tabular}
\end{table}
\section{Related Work}\label{relate-work}
\subsection{Integrating RIS into NOMA Transmissions}
Due to its capability of reconfiguring wireless channels and ease of deployment, RIS has been widely integrated into conventional wireless networks~\cite{Basar-tvtm2024}. The authors in~\cite{Chen-2023tcom} employed  RIS into  mobile edge computing (MEC) systems to assist workload offloading of edge users. By enhancing offloading channel conditions,  RIS provides edge users greater control flexibility in their offloading decisions. The authors in~\cite{Wang-2022twc} considered an RIS-assisted secure NOMA system, where  RIS enhances the SIC process for  authorized users while preventing signal leakage to eavesdroppers. Besides the single RIS system, multiple RISs can form cooperative beamforming or routing, further improving transmission performance~\cite{Huang-jsac2021}. The authors in~\cite{Liu-2023icc} explored a multi-RIS-aided multiple-input multiple-output (MIMO) system. A graph neural network (GNN)-based approach was proposed to map the complex multi-RIS scenario, and an RIS adaptive association mechanism was designed to cope with dynamic network conditions.
RIS can be further deployed on UAVs becoming ARIS, to provide more flexible services by exploiting UAVs' mobilities.
In~\cite{Zhang-2023twc}, ARIS dynamically plans its trajectory based on the current transmission requirements of ground users, which eliminates the need for deploying multiple terrestrial RISs to cover the entire wireless network. The authors in~\cite{Zhao-2025twc} explored a dual-mode UAV scheme, where the dual-mode UAVs are equipped with both  passive ARIS and  active antenna systems. By adaptively switching between the two modes, it can provide access and channel enhancement services for NOMA users. The above research demonstrates that the enhanced channel control flexibility enables more adaptable and enhanced NOMA transmissions. However, the NOMA decoding order needs to consider both channel conditions and users' traffic demands~\cite{Yang-2022tmc}. This motivates us to explore the JTAC scheme for improving NOMA transmission efficiency.
\subsection{Semantic Communication for Performance Enhancement}
Different from  bit-oriented transmission, semantic communication focuses on conveying the meanings of  raw data.
To achieve semantic communication, the authors in~\cite{yang-2023jsac} employed knowledge graphs by representing wireless entities (objects, places, or persons) as nodes and their relationships as edges. Then, the semantic information is extracted by filtering important graph structures. The authors in~\cite{Xie-2021tsp} proposed a deep learning-based semantic communication (DeepSC) framework, where a semantic encoder extracts semantic features from the original texts, while a channel encoder modulates the semantic information over the physical channel. By transmitting the lightweight semantic information, semantic communication can reshape network traffic and improve  spectral efficiency. Therefore,  semantic communication has been integrated into conventional wireless networks to break through their inherent performance limitations. The authors in~\cite{Cang-2024iotj} investigated a semantic-aware MEC system. By extracting the semantic features from the offloaded workload, it effectively reduces the transmission overhead of edge devices and improves the long-term energy saving performance of the system. The authors in~\cite{Li-2024cl} proposed a deep learning-based secure semantic communication (DeepSSC) framework that balances the trade-off between  security and reliability of the semantic transmission. By incorporating a security module into the semantic encoder,  DeepSSC introduces a new control dimension for secure communication compared to conventional physical layer security. The authors in~\cite{Mu-2023jsac} employed semantic communication in a two-user NOMA system, where the primary user transmits raw information while the secondary user transmits semantic information. The semantic extraction reshapes the secondary user's traffic demand, thus better improving the SIC process with the primary user. In contrast, this paper explores a more generalized multi-user scenario, where both channel reconfiguration and  traffic reshaping are integrated to enhance the NOMA transmission performance.
\subsection{Optimization in RIS-assisted Wireless Networks}
The performance maximization problems in RIS-assisted wireless networks are typically of high complexity and difficult to solve directly. This is because RIS's passive beamforming generally introduces high-dimensional variables, especially in scenarios involving  multiple RISs and dynamic network conditions~\cite{Liu-2024jsac}. Typically, the existing solution methods can roughly be categorized into  model-based optimization methods and  model-free learning methods. By exploiting the model information, the model-based optimization methods can effectively transform the complex problems into simplified standard forms that can be directly solved. The authors in~\cite{Wu-2020jsac} proposed a joint active and passive beamforming optimization method to minimize the system's energy consumption, where the original problem is decomposed into two subproblems and solved by using the AO and SDR methods.  In~\cite{Wu-2022twc}, they further explored the relationship between the number of beams and the number of users, proposing a dynamic beamforming optimization framework to balance the performance gain and the computational complexity. For adapting to dynamic scenarios, the authors in~\cite{Zhu-2022tcom} employed the Lyapunov optimization method to decompose the multi-slot problem into single-slot problems enabling efficient online solving. 
Some research employed model-free learning methods to adapt to complex networks, as they can learn the solutions from data or experience, such as~\cite{Hu-2024TVT} and~\cite{Ning-tvt2024}. The authors in~\cite{Hu-2024TVT} proposed a deep neural network (DNN)-based RIS optimization scheme, where a DNN is developed to map the channel conditions to the RIS's passive beamforming. The authors in~\cite{Ning-tvt2024} adopted a multi-agent DRL method in multi-RIS-assisted vehicular edge computing networks, where each RIS is treated as a DRL agent adapting its passive beamforming by interacting with the environment.
However, as the network complexity increases, the model-free learning methods become increasingly difficult to find the optimal solution. To tackle the above challenges, we propose a DRL-driven online optimization framework, where the optimization module focuses on improving the short-term transmission performance, while the DRL module controls the long-term network traffic. Moreover, the DRL module guides the optimization module to adaptively select the optimal option from traffic reshaping, channel reconfiguration, or NOMA decoding order assignment depending on the dynamic network status. This can avoid  iterative dependency and thus reduce the computational complexity significantly.

\section{System Model}\label{simultaneous-extraction-transmission}
\begin{figure}[t]
	\centering
	\includegraphics[width = 0.48\textwidth]{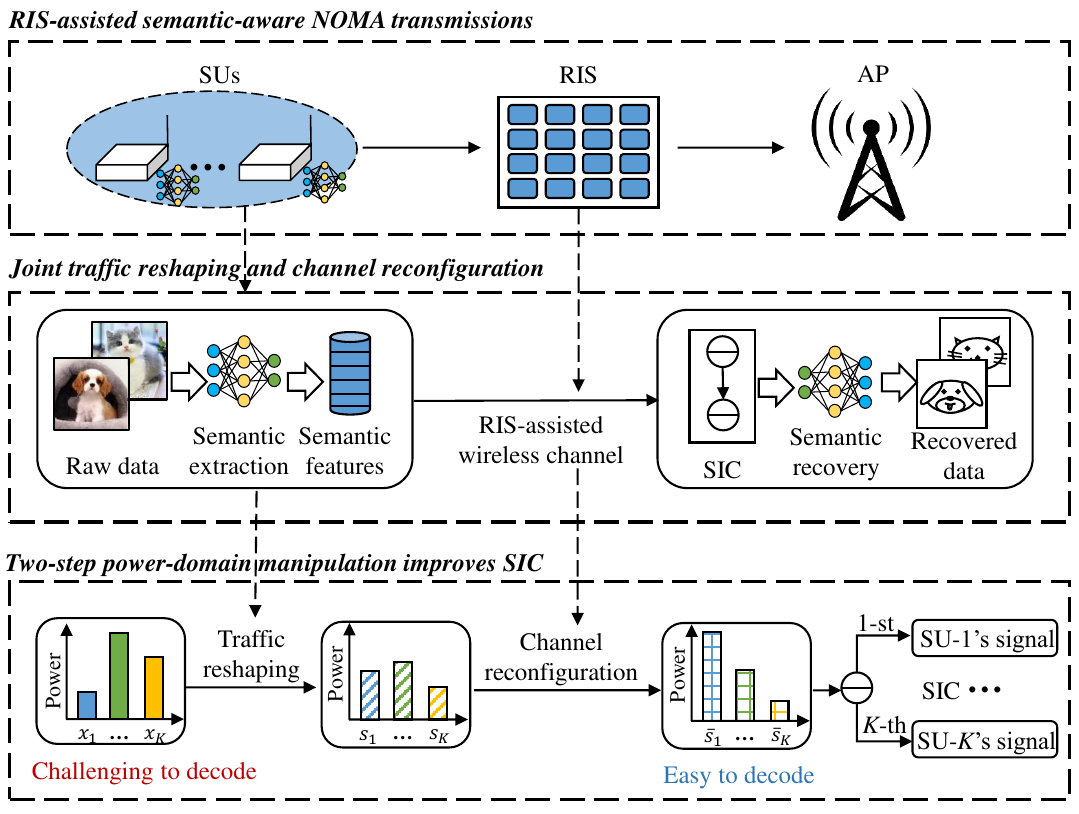}
	\caption{RIS-assisted semantic NOMA transmissions.}\label{system-model}
\end{figure}

We consider an RIS-assisted semantic-aware NOMA wireless network, where $K$ SUs periodically sense information from their surroundings and transmit it to the AP with the assistance of an RIS, as shown in Fig.~\ref{system-model}. Let $\mathcal{K}=\{1,\ldots,K\}$ denote the set of SUs, and the $k$-th SU is denoted as SU-$k$. We consider that all SUs and the AP are equipped with a single antenna. We employ the NOMA method for the SUs' simultaneous transmissions aiming to achieve higher spectral- and energy-efficiency compared to the conventional orthogonal multiple access (OMA) methods~\cite{Ding-2022twc}. Each SU is equipped with a semantic control unit enabling it to extract semantic information from  raw data~\cite{Cang-2024iotj}. The AP employs the SIC technique to decode each SU's semantic information from the superimposed signals one by one. After SIC decoding, the AP recovers each SU's raw data using a pre-trained semantic processor. The RIS is deployed between the AP and SUs to reconfigure the channel conditions via its passive beamforming. Note that the SUs' semantic extraction and RIS's passive beamforming jointly perform a two-step process to adjust the SUs' signal power to better meet their SIC requirements, which is expected to significantly enhance NOMA transmission efficiency.
\subsection{RIS-assisted Channel Model}
The RIS can induce passive beamforming to reconfigure the channel conditions between the AP and SUs. We consider that the RIS has $L$ reflecting elements, denoted by $\mathcal{L}=\{1,\ldots,L\}$. Let $t \in \mathcal{T}$ represent the time index, where $\mathcal{T}$ denotes the set of time slots. In each time slot, the RIS's passive beamforming strategy is indicated by $\boldsymbol{\theta}(t)=\{e^{j\phi_l(t)}\}_{l\in\mathcal{L},t\in\mathcal{T}}$, where $e^{j\phi_l(t)}$ is the phase-shift induced by the RIS's $l$-th reflecting element. The wireless channels from the AP to  SU-$k$ and the RIS as well as from the RIS to SU-$k$ are denoted by $h_{a,k}(t)$, ${\bf h}_{a,r}(t)$, and ${\bf h}_{r,k}(t)$, respectively. Hence, the equivalent channel $h_k(t)$ from the AP to  SU-$k$ is denoted as follows:
\begin{equation}\label{ris-channel}
h_k(t) = {\bf H}_k\boldsymbol{\theta}(t) + h_{a,k}(t),~k\in\mathcal{K},t\in\mathcal{T},
\end{equation}
where ${\bf H}_k(t) \triangleq {\bf h}_{a,r}(t)\text{diag}({\bf h}_{r,k}(t))$, and $\text{diag}({\bf h}_{r,k}(t))$ is the diagonal matrix with the diagonal vector ${\bf h}_{r,k}(t)$. Note that the equivalent channel $h_k(t)$ in~\eqref{ris-channel} can be reconfigured over time by dynamically designing the RIS's passive beamforming.
\subsection{Semantic NOMA Transmissions}
We consider that all SUs and the AP have  homogeneous semantic control units that can extract the semantic information at the SU side and then recover it at the AP~\cite{Mu-2023jsac2}. The semantic control units can shape the size of the extracted semantic information by controlling the extraction depth. {We adopt a linear semantic extraction model in this paper to describe the relationship among raw information, semantic information, and semantic extraction depth for simplicity.  This model can be extended to more complex forms depending on the specific semantic extraction methods.} Let $\rho_k(t)\in [\rho_{\min},1]$ denote  SU-$k$'s semantic extraction depth in the $t$-th time slot, where $\rho_{\min}$ is the minimum extraction threshold determined by the intended recovery accuracy~\cite{Yang-2023jasc}. {For brevity, we focus on the traffic-reshaping capability of semantic extraction while simplifying other performance aspects it may influence.} A smaller value of $\rho_k(t)$ indicates that SU-$k$ extracts smaller-size semantic information from the raw sensing data. After the semantic extraction process,  multiple SUs simultaneously transmit their semantic information to the AP by using the NOMA method. Then, the AP employs the SIC technique to decode all SUs' semantic information following a predefined decoding order.  We introduce a binary variable $\pi_{k,k'}(t)$ to denote the NOMA decoding order between  SU-$k$ and SU-$k'$. Specifically, let $\pi_{k,k'}(t)=1$ indicate that  SU-$k$'s signal is decoded earlier while considering the signal of SU-$k'$ as the interference. This results in the following decoding constraint:
\begin{subequations}\label{decoding-relation}
\begin{align}
&\pi_{k,k'}(t)+\pi_{k',k}(t) = 1 \text{ and }
\pi_{k,k'}(t)\in\{0,1\},\label{transform}\\
&r_k(t)\! < \!r_{k'}(t) \!+\! M (1-\pi_{k,k'}(t)),k\neq k',k\in\mathcal{K},t\in\mathcal{T},\label{added-rank}
\end{align}
\end{subequations}
where $\mathbf{r} = \{r_k(t)\}_{k\in\mathcal{K}}$ denotes the transmission priority  assigned for all SUs and $M$ is a sufficiently large constant to ensure acyclic decoding order~\cite{Huang-twc2024}. {Constraint~\eqref{added-rank} ensures that if $\pi_{k,k'}(t) = 1$ (i.e., SU-$k$ is decoded before SU-$k'$), then $r_k(t) < r_{k'}(t)$ holds, thus enforcing a consistent decoding order.} {Note that constraints~\eqref{decoding-relation} can also be extended to the imperfect SIC scenario by setting $\pi_{k,k'}(t)\in\{\beta,1\}$, where $\beta$ denotes the residual interference coefficient of imperfect SIC.}
Given the decoding order and  considering a normalized bandwidth, the semantic capacity $S_k(t)$ from SU-$k$ to the AP is represented as follows:
\begin{equation}\label{noma-throughtput}
\!S_k(t) \!= \!\tau\!\log_2\!\Big(\!1+\frac{|h_k(t)|^2p_k(t)}{\sum\limits_{k'\neq k,k'\in\mathcal{K}}\!\!\!\pi_{k,k'}(t)|h_{k'}(t)|^2 p_{k'}(t)\!+\!\sigma^2}\Big),
\end{equation}
where $p_k(t)$ and $\tau$ denote SU-$k$'s transmit power and the duration of each time slot, respectively. The background noise power at the AP is represented by $\sigma^2$. {When $\pi_{k,k'}(t) = 0$,  the interference power from SU-$k'$ can be perfectly removed by using the SIC, as shown in~\eqref{noma-throughtput}. In contrast, when $\pi_{k,k'}(t) = 1$, SU-$k$ treats the signal of SU-$k'$ as interference, and thus it is included in the interference term.}

In each time slot, SU-$k$ first senses raw data of size $\ell_k(t)$ from the environment and then extracts a portion of it into the semantic information for NOMA transmissions.
Let $B_{r,k}(t)$ denote  SU-$k$'s raw data buffer at the $t$-th time slot. Then, the dynamics of SU-$k$'s traffic demand are represented as follows:
\begin{equation}
B_{r,k}(t)\!=\!\max \Big\{ B_{r,k}(t-1)+\ell_k(t)-S_k(t)/\rho_k(t), 0\Big\},\label{Raw-Quene}
\end{equation}
{where \eqref{Raw-Quene} ensures that the system accurately tracks the data backlog and accounts for the effect of semantic extraction.}
Based on the Little's Law~\cite{Guizani-2010}, we impose the following time delay constraint to avoid excessive data backlog for the SUs.
\begin{equation}\label{delay-constraint}
\lim_{T\rightarrow\infty}\frac{1}{T}\sum_{t\in\mathcal{T}}\mathbb{E}[B_{r,k}(t)]\le B_{\max},~k\in\mathcal{K},
\end{equation}
where $\mathbb{E}[\cdot]$ represents the expectation operation, and $B_{\max}$ is the maximum buffer capacity determined by the maximum tolerable delay for the SUs. {The expectation in~\eqref{delay-constraint} is taken with respect to the stochastic variations in the channel states and SUs' data arrivals under the given control variables.} This implies that system control strategies should balance immediate performance and long-term system constraints.

Each SU's semantic capacity should not exceed its buffer capacity, while ensuring the successful decoding. This introduces the SUs' transmission requirements as follows:
\begin{equation}\label{su-qos}
B_{r,k}(t)\geq S_k(t)\geq S_{\min},~k\in\mathcal{K},t\in\mathcal{T},
\end{equation}
where $S_{\min}$ is the minimum capacity for successful decoding.
The sum semantic capacity $S_o(t)$ is represented as follows:
\begin{equation}\label{sum-throughtput}
S_o(t)\triangleq\sum_{k\in\mathcal{K}}S_k(t)\!=\!\tau\log_2\left(1\!+\!\frac{\sum_{k\in\mathcal{K}}|h_k(t)|^2p_k(t)}{\sigma^2} \right),
\end{equation}
{where the signal-to-interference-plus-noise ratio (SINR) terms in the sum rate expression progressively cancel out, leading to the simplified expression in~\eqref{sum-throughtput}~\cite{Zeng-icl2021}.}
Note that the NOMA decoding order does not affect the sum semantic capacity $S_o(t)$ in~\eqref{sum-throughtput}, but it directly impacts individual's semantic capacity $S_k(t)$ for each SU in~\eqref{noma-throughtput}.
Typically, given the RIS's passive beamforming, the decoding order can be determined by sorting all SUs' channel gains~\cite{Zheng-2020cl}. However, this method may not effectively satisfy constraints~\eqref{delay-constraint} and~\eqref{su-qos} for each SU. Therefore, the design of the NOMA decoding order should consider the specific states of each SU.
\subsection{Energy Budget for Semantic Control and Transmission}
The system's overall energy consumption depends on  both semantic control and transmission processes. The energy consumption in semantic control consists of both the SUs' semantic extraction and the AP's semantic recovery. Let $W_{e,k}(t) = a S_k(t)/\rho^{\alpha_e}_k(t)$ represent the additional computation load introduced by  SU-$k$'s semantic extraction, where $a>0$ and $\alpha_e>0$ denote the semantic extraction coefficients of the SUs' semantic processors. Similarly,  recovery of the semantic information at the AP incurs a computation load $W_{r,k}(t) = b S_k(t)/\rho^{\alpha_r}_k(t)$, where $b>0$ and $\alpha_r>0$ are the semantic recovery coefficients~\cite{Cang-2024iotj}. Note that $W_{e,k}(t)$ and $W_{r,k}(t)$ can also be modeled by alternative cost functions, which can be estimated by the specific semantic model. Thus, SU-$k$'s energy consumption for semantic control, denoted as $E_{s,k}(t)$, is represented as follows:
\begin{equation}\label{sem-energy_consumption}
E_{s,k}(t)=\kappa (f_k^2 W_{e,k}(t) + g^2 W_{r,k}(t)),~k\in\mathcal{K},t\in\mathcal{T},
\end{equation}
where $f_k$ and $g$ denote the computation capacities of SU-$k$ and the AP, respectively, and $\kappa$ is the energy efficiency coefficient that characterizes the relationship between the computational resource and the energy consumption~\cite{Bai-2021twc}.

The energy consumption in  SU-$k$'s NOMA transmission is characterized by $E_{t,k}(t) = \tau p_k(t)$. Combining the energy consumption from the semantic extraction and recovery, as well as the transmission processes, the system's overall energy consumption $E_o(t)$ is represented as follows:
\begin{equation}\label{overall-energy-consumption}
E_{o}(t)=\sum_{k\in\mathcal{K}}(E_{s,k}(t)+E_{t,k}(t)).
\end{equation}
Note that we cannot simply increase the depth of semantic extraction to reduce the transmission load, as this will require additional computational resources for both the extraction processes at the SUs and the recovery process at the AP.
\section{Energy Efficiency Maximization for RIS-assisted Semantic NOMA Transmissions}\label{pure-opti-method}
Defining $\eta(t) = S_o(t)/E_o(t)$ as the overall energy efficiency of the system, we aim to improve the long-term time-averaged energy efficiency $\lim_{T\rightarrow\infty}\frac{1}{T}\sum_{t\in\mathcal{T}}\eta(t)$. Note that different NOMA decoding orders place SUs to different interference conditions and thus affect both their semantic capacities and energy consumption. From an energy efficiency perspective, the SUs' semantic capacities should meet their traffic demands but do not have to be very high. Thus, it is important to design the SUs' NOMA decoding order to match with their dynamic traffic demands. Furthermore, the RIS can reconfigure the SUs' channel conditions and the semantic extraction can reshape the SUs' traffic demands. This implies that both factors can be jointly optimized to enhance the flexibility and adaptability of the NOMA transmissions. Therefore, we maximize  $\lim_{T\rightarrow\infty}\frac{1}{T}\sum_{t\in\mathcal{T}}\eta(t)$ by jointly optimizing the SUs' semantic extraction depth $\boldsymbol{\rho} = \{\rho_k(t)\}_{k\in\mathcal{K},t\in\mathcal{T}}$ and transmit power ${\bf p}=\{p_k(t)\}_{k\in\mathcal {K},t\in\mathcal{T}}$, the NOMA decoding order ${\boldsymbol \pi}=\{\pi_{k,k'}(t)\}_{k,k'\in\mathcal{K},t\in\mathcal{T}}$, and the RIS' passive beamforming strategy $\boldsymbol{\theta}=\{\boldsymbol{\theta}(t)\}_{t\in\mathcal{T}}$, as follows:
\begin{subequations}\label{energy-minimization}
\begin{align}
\max_{\boldsymbol{\rho},{\bf p},{\boldsymbol \pi},\boldsymbol{\theta}}&\lim_{T\rightarrow\infty}\frac{1}{T}\sum_{t\in\mathcal{T}}\eta(t), \\
\mathrm {s.t.}
~&~ \eqref{ris-channel}-\eqref{overall-energy-consumption},\\
~&~ \bm{\theta}(t)\in(0,2\pi)^L,0\le p_k(t)\le p_{\text{max}},\rho_{\min}\le\rho_k(t)\le1,\label{boundary-constraint}
\end{align}
\end{subequations}
where $p_{\max}$ is the SUs' maximum transmit power.
Problem~\eqref{energy-minimization} is intractable due to its mixed-integer nature, long-term optimization horizon, and the strong coupling among the control variables.  Although conventional model-based optimization methods can offer solutions to~\eqref{energy-minimization}, they generally rely on the perfect knowledge of the network states, which limits their capability to effectively optimize the long-term performance. Additionally, as the dimension of the control variables increases, the computational complexity of these optimization methods grows significantly, making them less practical for complex wireless systems.

To tackle these, we first consider a feasible case where each SU extracts all newly sensed raw data and transmits it within each time slot, i.e., $S_k(t)\ge\rho_k(t)\ell_k(t)$. {This allows problem~(10) to be decomposed into multiple per-slot subproblems for each time slot $t$, as follows:
\begin{subequations}\label{energy-minimization1}
\begin{align}
\max_{\boldsymbol{\rho},{\bf p},{\boldsymbol \pi},\boldsymbol{\theta}}~&~ \eta(t), \\
\mathrm {s.t.}
~&~ \eqref{ris-channel}-\eqref{noma-throughtput},\eqref{su-qos}-\eqref{overall-energy-consumption}, \text{ and } \eqref{boundary-constraint}  , \\
~&~ S_k(t)\ge\rho_k(t)\ell_k(t).
\end{align}
\end{subequations}
Since problem~\eqref{energy-minimization1} does not consider the long-term impact, this heuristic method provides only a suboptimal solution to problem~\eqref{energy-minimization}.}  Building on the insights gained from this approach, we further propose a DRL-driven online optimization framework that combines  both the model-free DRL and the model-based optimization methods, aiming to enhance the long-term performance while significantly reducing the computational complexity.
In the sequel, we focus on developing the optimization methods for the per-slot subproblem to verify the performance improvement by employing the JTAC scheme. For clarity, we omit the time index $t$ of all control variables.
\subsection{Penalty Relaxation for NOMA Decoding Order}\label{sectioniii-A}
We first optimize the NOMA decoding order while keeping the other control variables fixed. Then, the optimization of ${\boldsymbol \pi}$ becomes a feasibility check problem. A straightforward idea is to allocate a more favorable decoding order to ensure that the SUs' traffic demands are better satisfied.  This leads to the following subproblem by introducing the non-negative auxiliary variable ${\boldsymbol \mu} =\{\mu_k\}_{k\in\mathcal{K}}$:
\begin{subequations}\label{subproblem-decoding}
\begin{align}
\max_{{\boldsymbol \pi},{\boldsymbol \mu}}&~\sum_{k\in\mathcal{K}}~\mu_k, \\
\mathrm {s.t.}
~& \eqref{ris-channel}\text{ and }\eqref{decoding-relation},\\
~&\!\!\!\!\!\!\sum\limits_{k'\neq k,k'\in\mathcal{K}}\!\!\!\!\!\!\!\pi_{k,k'}|h_{k'}|^2 p_{k'}+\sigma^2+\mu_k\le\chi_{1,k},\forall k\in\mathcal{K},\label{interference-relax}
\end{align}
\end{subequations}
where $\chi_{1,k}\triangleq\frac{|h_k|^2p_k}{2^{\max\{\rho_k(t)\ell_k(t),S_{\min}\}/\tau}-1}$ and
the auxiliary variable $\mu_k$ can be explained as the additional interference that SU-$k$ can tolerate.  Then, the integer constraint~\eqref{transform} is relaxed to its continuous form as follows:
\begin{subequations}\label{continuous-relaxation}
\begin{align}
&\pi_{k,k'}-\pi_{k,k'}^2 = 0 \text{ and }\pi_{k,k'}\!+\!\pi_{k',k} - 1 =0,   \label{binary}    \\
&0 \leq \pi_{k,k'} \!\leq\! 1,~k\neq k',\forall k,k'\in\mathcal{K}.\label{plus-one}
\end{align}
\end{subequations}
{Incorporating strict equality constraint~\eqref{binary} directly into the original problem can severely restrict the feasible set, increasing the risk of infeasibility. To overcome this, we further employ the penalty method to approximate it by introducing a penalty parameter $\zeta$ and an auxiliary variable ${\boldsymbol\epsilon}=\{\epsilon _1,\epsilon_2\}$. By gradually increasing $\zeta$ during the optimization process, the solution can progressively approach that of the original problem.} Hence, problem~\eqref{subproblem-decoding} can be reformulated as follows:
\begin{subequations}\label{penalty-subproblem}
\begin{align}
\max_{{\boldsymbol \pi},{\boldsymbol \ell},{\boldsymbol\epsilon},\upsilon}&~\sum_{k\in\mathcal{K}}\mu_k -\zeta(\epsilon_1+\epsilon_2), \\
\mathrm {s.t.}
~& \eqref{ris-channel}, \eqref{added-rank}, \eqref{interference-relax}, \text{ and }\eqref{plus-one},\\
~&\sum_{k\in\mathcal{K}}\sum_{k\neq k',k'\in\mathcal{K}} \pi_{k,k'}-\pi_{k,k'}^2\le \epsilon_1,\label{epsilon-1}\\
~&\sum_{k\in\mathcal{K}}\sum_{k\neq k',k'\in\mathcal{K}}\pi_{k,k'}\!+\!\pi_{k',k} - 1\le \epsilon_2.\label{epsilon-2}
\end{align}
\end{subequations}
Constraint~\eqref{epsilon-1} can be further linearly approximated by the first-order Taylor expansion at given points $\pi_{k,k',0}$, as follows:
\begin{equation}\label{pi-taylor-expansion}
\sum_{k\in\mathcal{K}}\sum_{k\neq k',k'\in\mathcal{K}}\pi_{k,k'}+\pi_{k,k',0}^2-2\pi_{k,k'}\pi_{k,k',0}\le\epsilon_1.
\end{equation}
By substituting the linear approximation~\eqref{pi-taylor-expansion} into problem~\eqref{penalty-subproblem}, it becomes a convex optimization problem and can be directly solved by employing the standard convex solvers. {Since the penalty approach cannot strictly guarantee the binary nature, a rounding procedure is applied after convergence to obtain a feasible decoding order.}
\subsection{Beamforming Optimization for Channel Reconfiguration}\label{pass-opti}
The RIS's passive beamforming optimization aims to improve the SUs' semantic capacities according to their traffic demands.
Similar to problem~\eqref{penalty-subproblem}, by introducing the auxiliary matrices ${\bf G}_k = [{\bf H}_k, h_{a,k}]^H[{\bf H}_k, h_{a,k}]$ and ${\bf V} = [{\boldsymbol \theta};1][{\boldsymbol \theta};1]^H$, as well as the auxiliary variable ${\boldsymbol \upsilon}= \{\upsilon_k\}_{k\in\mathcal{K}}$, the optimization problem with respect to the RIS's passive beamforming can be reformulated as follows:
\begin{subequations}\label{RIS-subproblem}
\begin{align}
\max_{{\bf V},{\boldsymbol \upsilon} }&~\sum_{k\in\mathcal{K}}\upsilon_k,\\
\mathrm {s.t.}
~&~ \!\!\!\!\!\!\!\!\sum\limits_{k'\neq k,k'\in\mathcal{K}}\!\!\!\!\!\!\pi_{k,k'}\text{Tr}({\bf G}_{k'}{\bf V}) p_{k'}+\sigma^2+\upsilon_k\le\chi_{2,k},\label{re-sinr}\\
~&~{\bf V}\succeq0\text{ and }{\bf V}_{l,l}=1,~\forall l\in\{1,\ldots,L+1\},\label{semi-positive}\\
~&~\text{Rank}({\bf V})=1,\label{rank-one}
\end{align}
\end{subequations}
where $\chi_{2,k}\triangleq\frac{\text{Tr}({\bf G}_k{\bf V})p_k}{2^{\max\{\rho_k(t)\ell_k(t),S_{\min}\}/\tau}-1}$ and ${\bf V}_{l,l}$ denotes the $l$-th diagonal element of ${\bf V}$.
Subproblem~\eqref{RIS-subproblem} remains intractable due to the rank-one constraint. To solve this, we apply the sequential rank-one constraint relaxation (SROCR) method~\cite{Cao-2017european} by introducing a relaxation variable $\omega\in[0,1]$ as follows:
\begin{equation}\label{reformulate-rank-one}
{\bf e}^H{\bf V}{\bf e}\ge \omega\text{Tr}({\bf V}),
\end{equation}
where ${\bf e}$ is the eigenvector corresponding to the largest eigenvalue of ${\bf V}$. The main idea of SROCR method is to gradually tighten constraint~\eqref{reformulate-rank-one} to approximate~\eqref{rank-one} by sequentially increasing $\omega$ from $0$ to $1$. The vector ${\bf e}$ is updated in each iteration.
Substituting~\eqref{reformulate-rank-one} into~\eqref{RIS-subproblem}, it becomes a standard convex optimization that can be tackled directly.
\subsection{Semantic Extraction for Traffic Reshaping }
The SUs can adjust the semantic extraction depth to reshape their traffic demands as needed. The optimization problem with respect to $\boldsymbol{\rho}$ is reformulated as follows:
\begin{subequations}\label{semantic-rho}
\begin{align}
\min_{\boldsymbol{\rho}}&~\sum_{k\in\mathcal{K}}\frac{af_k^2 \ell_k}{\rho^{\alpha_e}_k}+ \frac{b g^2 \ell_k}{\rho^{\alpha_r}_k},\label{decreasing-rho} \\
\mathrm {s.t.}
~&~\eqref{noma-throughtput}, \rho_{\min}\le\rho_k\le1, \text{ and }
S_k\ge\rho_k\ell_k,\forall k\in\mathcal{K}.
\end{align}
\end{subequations}
Note that the objective function in~\eqref{decreasing-rho} is monotonically decreasing with respect to  $\boldsymbol{\rho}$. Thus, we can obtain its closed-form solution $\rho_k=\max\{S_k/\ell_k,\rho_{\min}\}$ by applying the boundary conditions of constraint~\eqref{su-qos}.

After obtaining the SUs' NOMA decoding order, the semantic extraction depth, and the RIS's passive beamforming, the SUs' channel conditions and traffic demands can be evaluated. Thus, the SUs' transmit power can be directly optimized using the similar method as in Section~\ref{pass-opti} and is omitted here for brevity. The detailed solution procedure is summarized in Algorithm~\ref{alg-jsap}. The subproblems are solved iteratively until the overall energy efficiency $\eta$ converges. {To ensure the convergence of Algorithm~\ref{alg-jsap}, we implement a heuristic convergence method. Specifically, we compare  $\eta$ across successive iterations and retain the solutions that yield a higher objective value. Meanwhile, the objective $\eta$ of problem~\eqref{energy-minimization1} has an upper bound due to its finite search space and continuous nature. Thus, Algorithm~\ref{alg-jsap} is guaranteed to converge to a stable value.} The computational complexity of Algorithm~\ref{alg-jsap} is estimated by $\mathcal{O}\big(K^{7}+(L+1)^{3.5} + K^{3.5} +K\big)I_{\max}$~\cite{Luo-2020spm}, where $I_{\max}$ is the maximum number of iterations.

\begin{algorithm}[t]
	\caption{Joint Optimization for NOMA Decoding Order, Semantic Extraction, and Passive Beamforming.}\label{alg-jsap}
	\begin{algorithmic}[1]
        \State Initialize the iterative index $i=0$, parameters $\zeta^{(i)}$ and $\omega^{(i)}$, as well as decoding order ${\boldsymbol \pi}^{(i)}$. Set the overall energy efficiency $\eta^{(i)}$ and convergence threshold $\epsilon>0$.
        \State \textbf{repeat}
        \State \hspace{3mm}$i = i+1$
        \State \hspace{3mm}Update ${\boldsymbol \pi}^{(i)}$ by solving subproblem~\eqref{penalty-subproblem}
        \State \hspace{3mm}Update $\boldsymbol{\theta}^{(i)}$ by solving subproblem~\eqref{RIS-subproblem}
        \State \hspace{3mm}Update       $\boldsymbol{\rho}^{(i)}$ by solving subproblem~\eqref{semantic-rho}
        \State \hspace{3mm}Update $\eta^{(i)}$ by employing~\eqref{sum-throughtput} and~\eqref{overall-energy-consumption}
        \State \textbf{until} $\eta^{(i)}-\eta^{(i-1)}\leq\epsilon$
	\end{algorithmic}
\end{algorithm}

\subsection{Validation and Insight of the JTAC Scheme}\label{val-con-jtac}
\begin{figure}[t]
	\centering
	\includegraphics[width = 0.45\textwidth]{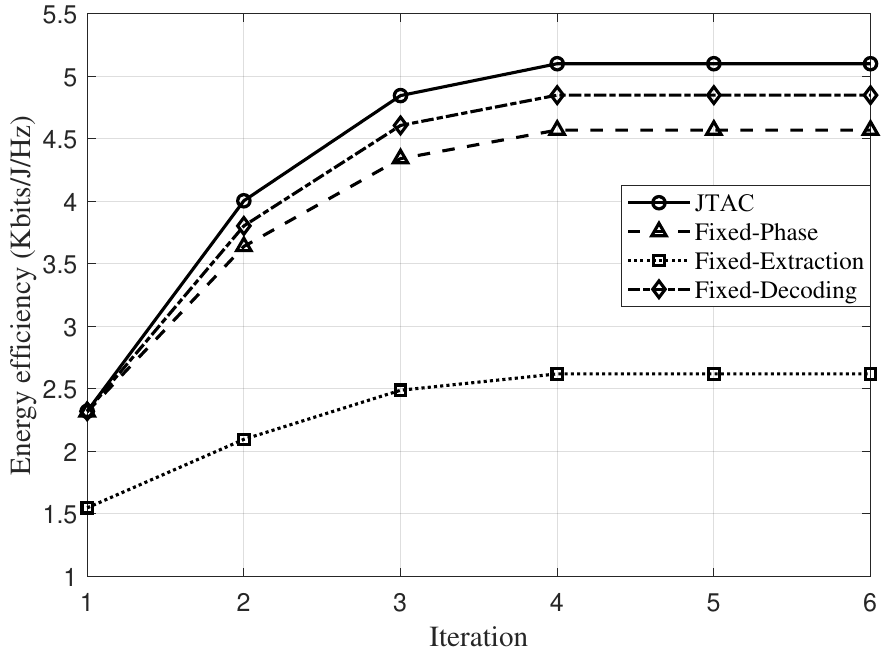}
	\caption{Experimental verification of the JTAC scheme.}\label{test-for-jtac}
\end{figure}

We verify the effectiveness of the proposed JTAC scheme, as shown in Fig.~\ref{test-for-jtac}. {The default parameters are set as follows: $K=3$, $p_{\max} = 40$ dBm, $\tau =1$, $\rho_{\min} = 0.2$, $\ell_k=1$ Kbits, $\kappa = 10^{-21}$, $a =100$, $b = 200$, $\alpha_e =4$, $\alpha_r =2$, $f_k = 5\times10^{8}$ cycles/s, $g=10^9$ cycles/s, $\epsilon = 10^{-3}$, and $\sigma^2 = -90$ dBm.} We apply Algorithm~\ref{alg-jsap} and examine the energy efficiency performance of the JTAC scheme compared with three benchmark schemes: Fixed-Phase, Fixed-Extraction, and Fixed-Decoding, where the RIS's passive beamforming, the SUs' semantic extraction, and NOMA decoding order remain static during the optimization, respectively. It is observed that all schemes rapidly reach the stable value, verifying the convergence of Algorithm~\ref{alg-jsap}. {Moreover, the JTAC scheme achieves the highest energy efficiency among all benchmarks that lack optimization of one control variable. This demonstrates that jointly optimizing all control variables offers significant flexibility and performance gains for NOMA transmissions.} As such, the SUs can adapt their transmission strategies more effectively and thus significantly improve the energy efficiency.
\section{Deferrable Semantic Extraction Scheme with DRL-driven Online Optimization}\label{deferred-PGOO}
Note that Algorithm~\ref{alg-jsap} focuses on improving the multiple SUs' transmission performance in static single slot scenarios. However, it will face limitations in a dynamic multi-slot scenarios. This is because the alternating iteration process of Algorithm~\ref{alg-jsap} can introduce significant computational overhead, which hinders its effectiveness for the real-time execution. Additionally, since the SUs' traffic demands and channel conditions vary across time slots, it becomes inefficient for all SUs to participate in the NOMA transmissions in every time slot. 
Instead, the SUs can distribute their raw data over multiple time slots for semantic extraction and defer the transmission until more favorable channel conditions appear. Therefore, in this section, we aim to design a deferrable semantic extraction scheme tailored for the long-term transmission scenarios. Furthermore, we propose a lightweight online optimization method to reduce the computational complexity and improve the learning efficiency.
\subsection{Deferrable Semantic Extraction and Adaptive Transmission}
Different from real-time semantic extraction in Section~\ref{simultaneous-extraction-transmission},  SU-$k$ in deferrable semantic extraction scheme can choose different sizes $D_k(t)$ and depths $\rho_k(t)$ for the semantic extraction across multiple slots and delay the semantic transmissions until appropriate network conditions appear. The deferrable semantic extraction is expected to improve the control flexibility and thus improve the SUs' transmissions significantly.

Let binary variable $\psi_k(t)$ denote  SU-$k$'s transmission strategy, where $\psi_k(t)=1$ indicates that SU-$k$ joins NOMA transmissions at the $t$-th time slot. Given $\psi_k(t)$,  SU-$k$'s semantic capacity $\widehat{S}_k(t)$ is reformulated as follows:
\begin{equation}\label{sem-cap-long}
\widehat{S}_k=\tau\log_2\Big(1+\frac{\psi_k|h_k|^2p_k}{\sum\limits_{k'\neq k,k'\in\mathcal{K}}\psi_{k'}\pi_{k,k'}|h_{k'}|^2 p_{k'}+\sigma^2}\Big).
\end{equation}
Under the deferrable semantic extraction,  SU-$k$ has  the following two buffer dynamics (i.e., $\widehat{B}_{r,k}(t)$ and $\widehat{B}_{s,k}(t)$) for the raw data and semantic information, respectively:
\begin{subequations}\label{semantic-subproblem}
\begin{align}
&\!\!\!\widehat{B}_{r,k}(t)\!=\!\max\! \Big\{ \widehat{B}_{r,k}(t-1)+\ell_{k}(t)-D_{k}(t), 0\Big\},\\
&\!\!\!\widehat{B}_{s,k}(t)\!=\!\max\! \Big\{ \widehat{B}_{s,k}(t-1)+\rho_k(t)D_{k}(t)-\widehat{S}_k(t), 0\Big\}.
\end{align}
\end{subequations}
The raw data and the extracted semantic information are stored in separate buffers, which helps avoid repetitive semantic extraction of the raw data.
Given~\eqref{semantic-subproblem}, we can update the time delay constraint as follows:
\begin{equation}\label{delay-constraint2}
\lim_{T\rightarrow\infty}\frac{1}{T}\sum_{t\in\mathcal{T}}\mathbb{E}[\widehat{B}_{r,k}+\widehat{B}_{s,k}]\le \widehat{B}_{\max},~k\in\mathcal{K},
\end{equation}
where $\widehat{B}_{\max}$ can be pre-determined based on the SUs' delay requirement, similar to~\eqref{delay-constraint}.

By defining $\boldsymbol{\psi} = \{\psi_k\}_{k\in\mathcal{K}}$ and ${\bf D}=\{D_k\}_{k\in\mathcal{K}}$, the long-term average energy efficiency maximization problem can be reformulated as follows:
\begin{subequations}\label{long-term-opti}
\begin{align}
\max_{\boldsymbol{\rho},{\bf p},{\boldsymbol \pi},\boldsymbol{\theta},{\bf D},\boldsymbol{\psi}}&\lim_{T\rightarrow\infty}\frac{1}{T}\sum_{t\in\mathcal{T}}\widehat{\eta} , \notag\\
\mathrm {s.t.}
~& \eqref{ris-channel}-\eqref{decoding-relation},\eqref{sem-energy_consumption}-\eqref{overall-energy-consumption},\text{ and } \eqref{sem-cap-long}-\eqref{delay-constraint2},\\
&\widehat{B}_{s,k}\geq \widehat{S}_k\geq S_{\min},~\forall k\in\mathcal{K},\\
~& \psi_k=\{0,1\},~\forall k\in\mathcal{K},
\end{align}
\end{subequations}
where $\widehat{\eta} = 1/E_o\sum_{k\in\mathcal{K}} \widehat{S}_k/\rho_k$. 
A straightforward solution is to decompose problem~\eqref{long-term-opti} into multiple per-slot subproblems and solve each subproblem using  Algorithm~\ref{alg-jsap}, as proposed in Section~\ref{pure-opti-method}. However, {directly decomposing the long-term problem into per-slot subproblems may overlook inter-slot dependencies and the long-term impact of short-term decisions.} Another challenge lies in that the alternating iteration of Algorithm~\ref{alg-jsap} will introduce high computational complexity in dynamic scenarios, thus making it challenging for online execution. {To tackle these challenges, we propose a DRL-driven online optimization framework that leverages the DRL's capability to optimize long-term decision-making by learning from past experiences. The DRL module adapts long-term-impacting variables through continuous interaction with the environment. Once these variables are determined, the model-based optimization methods are employed to improve the per-slot transmission efficiency.}

\subsection{DRL-driven Optimization for Online Execution}
\begin{figure}[t]
	\centering
	\includegraphics[width = 0.45\textwidth]{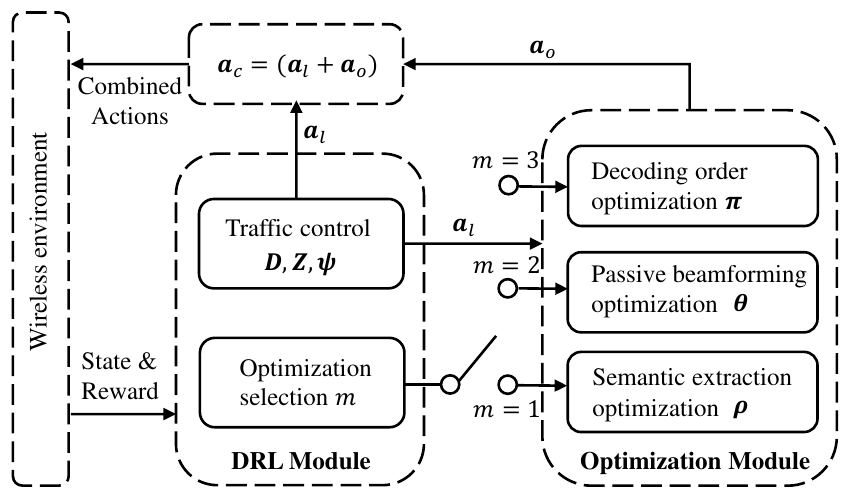}
	\caption{DRL-driven online optimization framework.}\label{fig-drl-opti}
\end{figure}
The first limitation of applying Algorithm~\ref{alg-jsap} to long-term problem~\eqref{long-term-opti} lies in that we do not know how much semantic information the SUs need to extract in each time slot and the timing for them to join the NOMA transmissions. If the above information were available, Algorithm~\ref{alg-jsap} could be readily adapted for online execution in each time slot. Since DRL has the capability to optimize long-term decision-making by learning from past experiences, this motivates us to develop the DRL-driven online optimization framework, as shown in Fig.~\ref{fig-drl-opti}. The DRL agent adaptively determines the size of raw data ${\bf D}$ to be extracted, the size of the semantic data ${\bf Z}=\{Z_k\}_{k\in\mathcal{K}}$ to be transmitted, and the transmission strategy $\boldsymbol{\psi}$. Thus, we can reformulate SUs' traffic demands as follows:
\begin{equation}\label{su-qos-long}
\widehat{S}_k\geq \psi_k\max\{Z_k, S_{\min}\} ,~k\in\mathcal{K}.
\end{equation}
Given $\{{\bf D}, {\bf Z},\boldsymbol{\psi}\}$ and introducing~\eqref{su-qos-long} into problem~\eqref{long-term-opti}, it can be easily transformed into multiple independent per-slot problems, which can be directly solved using Algorithm~\ref{alg-jsap}.

To further reduce the computational complexity of Algorithm~\ref{alg-jsap}, we aim to design lightweight algorithms enabling efficient online execution in dynamic scenarios. To this end, we first delve into three subproblems in Algorithm~\ref{alg-jsap} to inspire the design. Specifically, the RIS's passive beamforming optimization can reconfigure the SUs' channel conditions, while the semantic extraction optimization can adapt to variations in SUs' traffic demands. When both factors change significantly, the NOMA decoding order optimization can reallocate all SUs' transmission priorities. It is observed that each optimization serves a specific function, which motivates us to adaptively choose one preferable optimization depending on the current network status. We envision that the optimization module executed online does not need to be precise but requires low complexity in each time slot. The performance loss can be compensated by  DRL's learning capability. Therefore, we employ DRL to adaptively select one optimization to avoid redundant iterations during operation. Let $m=\{1,2,3\}$ define the optimization selection strategy corresponding to the semantic extraction, the RIS's passive beamforming, and the NOMA decoding order optimizations, respectively.

We employ the proximal policy optimization (PPO) algorithm in the proposed DRL-driven online optimization framework, since its advantage function and clipping mechanism can effectively improve the convergence and learning efficiency~\cite{Gong-2023twc}. As shown in Fig.~\ref{fig-drl-opti},  the PPO-driven online optimization framework (denoted as PDOO) consists of the DRL module and optimization module. The DRL module is characterized by the tuple ($\mathcal{S}, (\mathcal{A}_{l},\mathcal{A}_{o}),\mathcal{R}$). The state space $s\in\mathcal{S}$ includes the SUs' channel conditions and traffic demands. All control variables are divided into two action spaces, governed by the DRL and optimization modules, i.e., $\mathbf{a}_l=\{{\bf D},{\bf Z},\boldsymbol{\psi},m\}\in\mathcal{A}_{l}$ and $\mathbf{a}_o=\{\boldsymbol{\rho},{\boldsymbol \pi},\boldsymbol{\theta}\}\in\mathcal{A}_{o}$, respectively. Based on the current network status, the DRL agent adjusts $\mathbf{a}_l=\{{\bf D},{\bf Z},\boldsymbol{\psi},m\}$ to control the SUs' traffic demands and obtain the optimization selection strategy. Given $\mathbf{a}_l$, the optimization module solves the selected optimization and updates the corresponding control variable. Combining the actions from both the optimization and DRL modules, i.e., $\mathbf{a}_c=(\mathbf{a}_l+\mathbf{a}_o)$, and executing them within the wireless environment, we can observe the feedback reward $r$ to evaluate the goodness of the state-action pair.

To maximize the overall energy efficiency and satisfy the constraints, we define the reward function $r\in\mathcal{R}$ as follows:
\begin{equation}\label{reward-function}
r=\widehat{\eta}-\sum_{k\in\mathcal{K}}\lambda_k\Big(\frac{1}{H}\sum_{t\in\mathcal{H}_t}\big(\widehat{B}_{r,k}+\widehat{B}_{s,k}\big)-\widehat{B}_{\max}\Big),
\end{equation}
where $\lambda_k$ is the penalty parameter and  $\mathcal{H}_t\triangleq\{t-H+1,\ldots,t\}$ denotes the sliding window over the past $H$ time slots. The second term in~\eqref{reward-function} ensures the delay constraint in~\eqref{delay-constraint2}. The detailed procedures of the PDOO algorithm are outlined in  Algorithm~\ref{alg-pgoo}. The PPO agent first learns the long-term traffic control variables and optimization selection strategy as shown in lines~\ref{observe-state}-\ref{adapt-actions}. Given the guidance from the PPO, the optimization module optimizes the corresponding subproblem as in line~\ref{optimize-subproblem}. In lines~\ref{com-act}-\ref{update-networks}, the PPO agent executes the joint action $\mathbf{a}_{c}=\{\mathbf{a}_{l},\mathbf{a}_{o}\}$ from both modules in the environment. Then, it observes the feedback reward $r$ and transitions to the next state ${\bf s}'$. All interactions are stored in the experience replay buffer for the subsequent training of the DNN.
{In each interaction, the overall computational complexity is characterized as $C_d +C_m$, where $C_d$ and $C_m$ denote the complexities of the DRL module and the optimization module, respectively. Given the optimization selection strategy $m$, the complexity of optimization module is calculated as: $C_1 = \mathcal{O}(K)$, $C_2 = \mathcal{O}((L+1)^{3.5})$, and $C_3 =\mathcal{O}(K^7)$, respectively.   Let $ n_{a,f}$ and $n_{c,f}$ represent the number of neurons in the $f$-th layer of the actor and critic networks, respectively. The complexity of the DRL module can be expressed as $C_d = \mathcal{O}\Big(\sum_{f=0}^{F_a-1}n_{a,f}n_{a,f+1} + \sum_{f=0}^{F_c-1}n_{c,f}n_{c,f+1}\Big)$~\cite{Guo-ton2023}, where $F_a$ and $F_c$ denote the layer number of the actor and critic networks, respectively.}

\begin{algorithm}[t]
	\caption{PDOO for Traffic Reshaping and Channel Reconfiguration in RIS-assisted Semantic NOMA Transmissions.}\label{alg-pgoo}
	\begin{algorithmic}[1]
        \State Initialize the DNN parameters and all control variables.
        \Statex \textbf{\% Long-term guidance by the PPO}
        \State \hspace{3mm} Identify the system state $s$\label{observe-state}
        \State \hspace{3mm} PPO agent adapts long-term variables $\{ {\bf Q},{\bf Z},\boldsymbol{\psi}\}$ and \label{adapt-actions}
        \Statex \hspace{4mm}optimization module selection $m$
        \Statex \textbf{\% Online execution by the selected optimization}
        \State \hspace{3mm} Given $m$, the optimization module selects the \label{optimize-subproblem}
        \Statex \hspace{4mm}corresponding control variable to optimize
        \State \hspace{3mm}  Execute the combined action \label{com-act} $\mathbf{a}_{c}=\{\mathbf{a}_{l},\mathbf{a}_{o}\}$
        \State \hspace{3mm} Observe the reward function $r$\label{observe-reward}
        \State \hspace{3mm} Record the transition to the next state ${\bf s}'$ \label{next-state}
        \State \hspace{3mm} Store the transition sample $\{{\bf s},\mathbf{a}_{c}, r, {\bf s}'\}$\label{store-transition}
        \State \hspace{3mm} Update the parameters of the actor- and critic-networks\label{update-networks}
	\end{algorithmic}
\end{algorithm}

\subsection{Lightweight Design for Optimization Module}\label{lightweight-pgoo}
The proposed PDOO algorithm employs the DRL module to dynamically select one specific subproblem for online execution. This avoids the alternating iterations among the subproblems,  thus significantly reducing overall computational complexity. However, the optimizations of the NOMA decoding order and the RIS's passive beamforming may introduce considerable computational overhead, i.e., $\mathcal{O}(K^7)$ and $\mathcal{O}((L+1)^{3.5})$, especially as the dimensionality of control variables increases. To further enhance the learning efficiency of PDOO framework, we aim to design lightweight method for the optimization module.

The first design is to propose the heuristic method to optimize the NOMA decoding order.
In every learning interaction, the objective of the NOMA decoding order optimization is to minimize the SUs' transmission energy, as their traffic demands have already been determined by the PPO.
Let SU-$i$ and  SU-$j$, where $\forall i,j\in\mathcal{K}$, be two adjacent decoded SUs simultaneously joining the NOMA transmissions. Considering that SU-$i$ decodes before SU-$j$, we have their SINR requirements as follows:
\begin{equation}\label{ana-tra-demand}
\frac{p_i|h_i|^2}{p_j|h_j|^2+I_o+\sigma^2}\ge\omega_i \text{ and }\frac{p_j|h_j|^2}{I_o+\sigma^2}\ge\omega_j,
\end{equation}
where $\omega_i = 2^{\max\{Z_k(t),S_{\min}\}/\tau} - 1$. The parameter $I_o$ denotes the interference  from the other SUs that are decoded later than SU-$i$ and  SU-$j$.
Based on~\eqref{ana-tra-demand}, we calculate the lower-bound of the sum transmit power of  SU-$i$ and  SU-$j$ as follows:
\begin{equation}\label{lower-bound-power}
p_i+p_j \ge \frac{\omega_i(\sigma^2+I_o)}{|h_i|^2} + \frac{\omega_j(\sigma^2+I_o)}{|h_j|^2}+ \frac{\omega_i\omega_j(\sigma^2+I_o)}{|h_i|^2}.
\end{equation}

It is observed that only the third term of~\eqref{lower-bound-power} varies with the decoding order, i.e.,  $h_i$ of the first decoded SU-$i$. If  SU-$i$ has a better channel gain, the third term becomes smaller. This allows us to heuristically determine the decoding order by sorting the channel gains in descending order. Note that this result differs from the conclusion in~\cite{Yang-2022tmc}, where the decoding order is determined based on both traffic demands and channel conditions. This is because the SUs' traffic demands have already been controlled by the PPO,  thus eliminating its impact on the decoding order optimization. Then, we aim to design a heuristical method to  reduce the complexity of the RIS's passive beamforming optimization. Instead of using the SDR method for the RIS's passive beamforming optimization, we can align the RIS's phase shift $\boldsymbol{\theta}$ with the SU that currently has the highest traffic demand. By adopting the heuristical methods above, the computational complexities are significantly reduced to linear complexities, i.e., $\mathcal{O}(K)$ and $\mathcal{O}(L)$. It is expected that the above designs can significantly reduce the training time while preserving attainable performance.
\section{Numerical Results}\label{results}
\begin{table}[t]
\caption{Parameter settings in the simulations.} \label{para_settings}\normalsize
	\centering
{
	\begin{tabular}{|l|l|}
		\hline
        Parameters&Settings\\
        \hline
        Number of SUs& $K=3$ \\
        Data arrival rate  & $\ell_k=1$ Kbit/s \\
        Duration of each time slot& $\tau$ = 1\\
        Maximum buffer capacity& $B_{\max} = 3$ Kbits\\
        SUs' maximum transmit power& $p_{\max}=40$ dBm\\
        Minimum extraction threshold& $\rho_{\min} = 0.2$\\
        Minimum transmission threshold& $S_{\min}= 0.1$ Kbit/s\\
		Energy efficiency coefficient & $\kappa =10^{-21}$ \\
		Size of the RIS's elements& $L=70$\\
        Background noise power & $\sigma= -90$ dBm\\
		Semantic extraction coefficients& $a = 100$ and $a_e = 4$\\
        Semantic recovery coefficients & $b= 100$ and $a_r = 2$ \\
        SUs' computation capacities& $f_k = 5\times10^8$  cycles/s\\
        AP's computation capacities& $g =10^9$  cycles/s\\
        Convergence threshold of Alg.~1& $\epsilon = 10^{-3}$\\
        Actor's learning rate& $2\times10^{-4}$\\
        Critic's learning rate&$2\times10^{-4}$\\
        Reward discount factor &$0.99$\\
        \hline
	\end{tabular}}
\end{table}
We explore the proposed JTAC scheme in the RIS-assisted semantic NOMA wireless networks. We also evaluate the performance enhancement of the PDOO algorithm for JTAC compared to benchmarks. In the simulations, we consider that the AP is placed at the coordinate origin and the RIS is placed at $(5,0)$. The SUs are randomly distributed around $(7,3)$.  {We consider Rician fading for all channels and the detailed channel parameters are set according to~\cite{Wu-2019twc}. The actor and critic networks are initialized using Xavier initialization. PPO updates are performed every $128$ steps of collected trajectories. The random number generators in Python's random module, NumPy, and PyTorch are all seeded with $42$. We employ MOSEK as solvers in Section~\ref{sectioniii-A} and~\ref{pass-opti}, with a stopping criterion of $10^{-3}$ per subproblem. SUs' data rates are generated per time slot following a Gaussian distribution with a default mean of $1$ Kbit/s.} The default parameters are listed in Table~\ref{para_settings}, following the similar settings in~\cite{Cang-2024iotj}.
\begin{figure}[t]
	\centering
    \subfloat[Convergence of the PDOO framework.]{\includegraphics[width=0.45\textwidth]{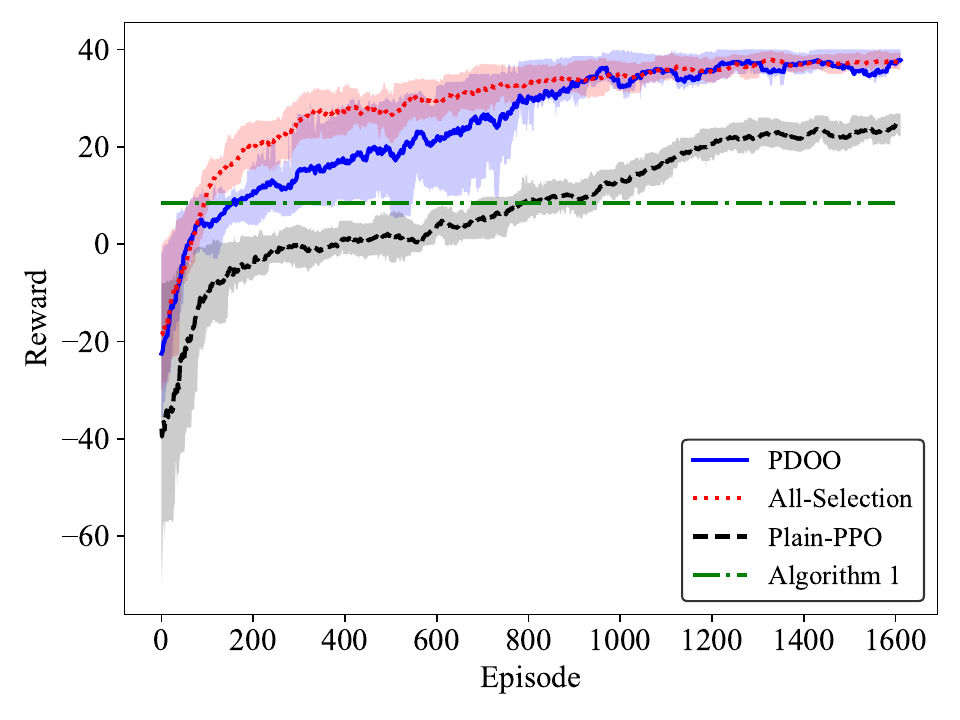}}\\
	\subfloat[Convergence with varying learning rates.]{\includegraphics[width=0.45\textwidth]{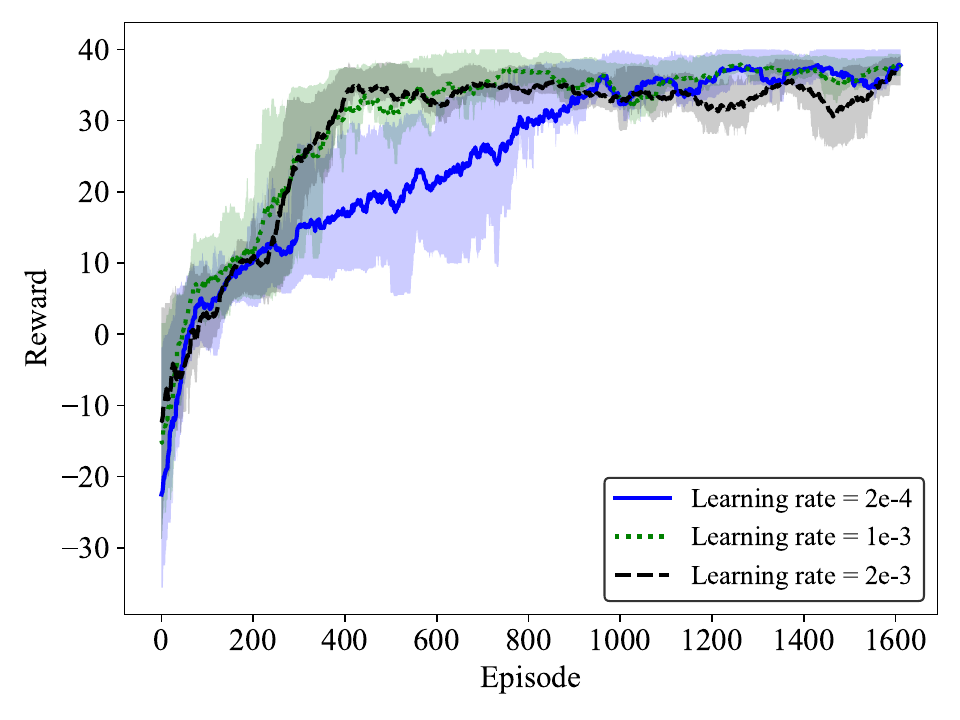}}
	\caption{PDOO improves the optimization efficiency.}\label{PGOO-convergence}
\end{figure}
\subsection{Convergence of PDOO Framework}
In the PDOO framework, the PDOO agent adaptively selects one subproblem in the optimization module to optimize depending on the current network status. We aim to validate that the PDOO framework can effectively enhance learning efficiency while maintaining satisfactory performance. We compare the PDOO with three benchmark methods: ALL-Selection, Plain-PPO, and Algorithm~\ref{alg-jsap}. In ALL-Selection, the PPO agent only controls the traffic demands while all subproblems are selected and alternately optimized in each PPO learning step, similar to the algorithm framework in~\cite{Zhao-2025twc}. As such, the optimization module in ALL-Selection can achieve more precise solution compared to the PDOO. In Plain-PPO, all control variables are directly learned by the PPO agent. In Algorithm~\ref{alg-jsap}, it is assumed that all newly sensed data of the SUs is extracted and transmitted in each time slot. In this case, all control variables can be optimized by  Algorithm~\ref{alg-jsap}. {Specifically, ALL-Selection lacks the mode selection strategy, Plain-PPO lacks the optimization module, and Algorithm~\ref{alg-jsap} lacks the DRL module. These designs enables an effective evaluation of the contribution of each module in the proposed PDOO framework.}

In Fig.~\ref{PGOO-convergence}(a), it is observed that the PDOO achieves superior learning performance compared to Plain-PPO and Algorithm~\ref{alg-jsap}. The reasons are as follows:
The PDOO framework employs the model-based optimization methods to exploit the model information. This allows the PPO agent to better understand the wireless environment, thus enabling it to more efficiently learn in complex environment. Moreover, the optimization module significantly reduces the action space, enabling the PDOO agent to learn the actions  more effectively compared to that in the model-free Plain-PPO.
Although Algorithm~\ref{alg-jsap} can obtain more precise solutions in each time slot, it focuses only on per-slot performance enhancement. Hence, it struggles to schedule the resources across multiple slots, thus leading to lower energy efficiency performance in long-term scenarios.
It is interesting to observe that PDOO learns nearly the same reward performance as ALL-Selection. However, the average training time of PDOO is significantly reduced to $0.1857$ s/step compared to $1.0273$ s/step in ALL-Selection.  This validates that we only need to select and solve the most efficient subproblem in each decision-making step, while the reward performance is still ensured by the DRL's long-term learning capability. We also evaluate the impact of the learning rate on the PDOO's learning performance, as shown in Fig.~\ref{PGOO-convergence}(b). The PDOO efficiently converges to the stable values under different learning rates. This is because part of the control variables is assigned to the optimization module. The PPO only needs to explore a reduced space, and thus effectively improves its robustness.
\subsection{Energy Efficiency Performance with PDOO Framework}
\begin{figure}[t]
	\centering
	\subfloat[Changes in RIS size.]{\includegraphics[width=0.22\textwidth]{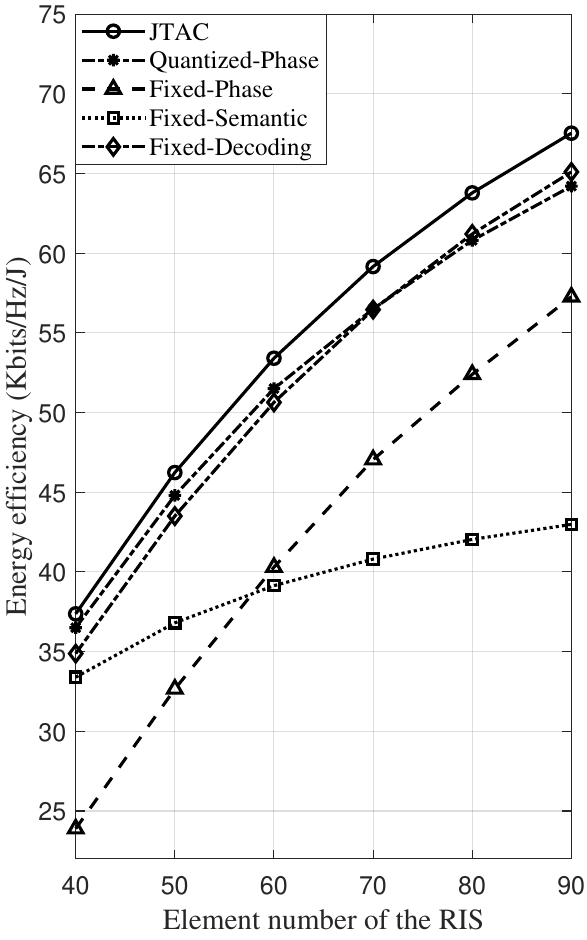}}
	\subfloat[Changes in RIS location.]{\includegraphics[width=0.22\textwidth]{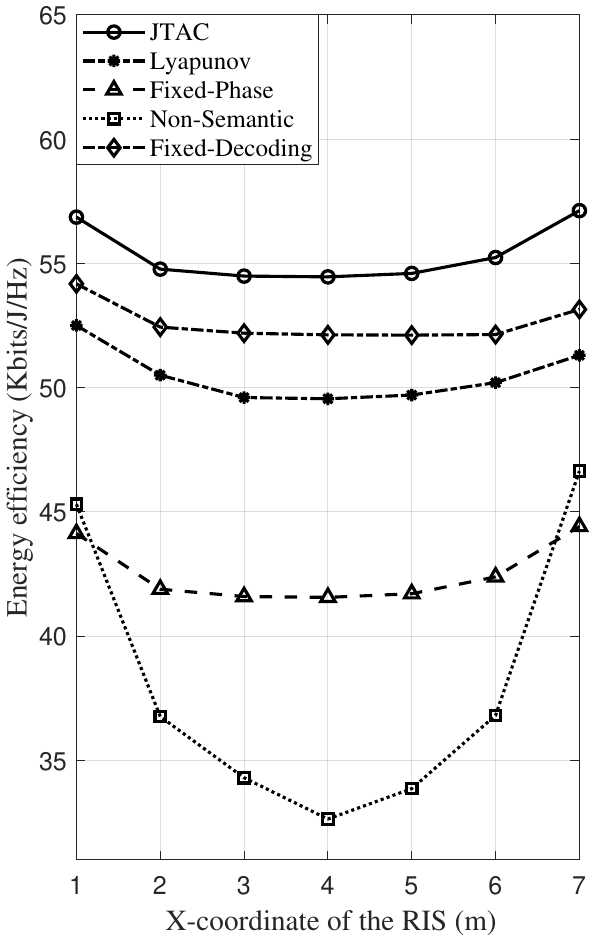}}
	\caption{Energy efficiency under different RIS settings.}\label{fig-ris-settings}
\end{figure}
We evaluate the energy efficiency of the proposed PDOO method as we increase the size of the RIS. To study the impact of the optimization module on the PDOO framework, we compare different transmission schemes adopted in the optimization module, i.e., JTAC, Fixed-Phase, Fixed-Semantic, and Fixed-Decoding, as described in Section~\ref{val-con-jtac}.  {We also study the impact of quantized RIS (denoted as Quantized-Phase), assuming $4$ discrete phase shift levels. The continuous phase values are mapped to the nearest discrete levels to obtain the quantized phase shifts.} As shown in Fig.~\ref{fig-ris-settings}(a), the energy efficiency of all schemes increases as the size of the RIS increases. The JTAC scheme achieves  higher energy efficiency than all the benchmark schemes. {As $L$ increases, the performance gap between JTAC and Quantized-Phase also grows due to the accumulation of quantization errors.} Meanwhile, a large size of the RIS leads to the better channel conditions. As such, the JTAC scheme achieves a higher energy efficiency by reducing the energy for semantic extraction while allocating more energy to the transmission process.  This observation is further validated by the performance of the Fixed-Semantic scheme. Due to the fixed semantic extraction strategy, it has limited capability to balance the energy consumption between the semantic extraction and transmission. Thus, even with the enhanced channel conditions, the energy efficiency of the Fixed-Semantic scheme shows only a slower improvement compared to other schemes.
\begin{figure}[t]
	\centering
	\includegraphics[width = 0.45\textwidth]{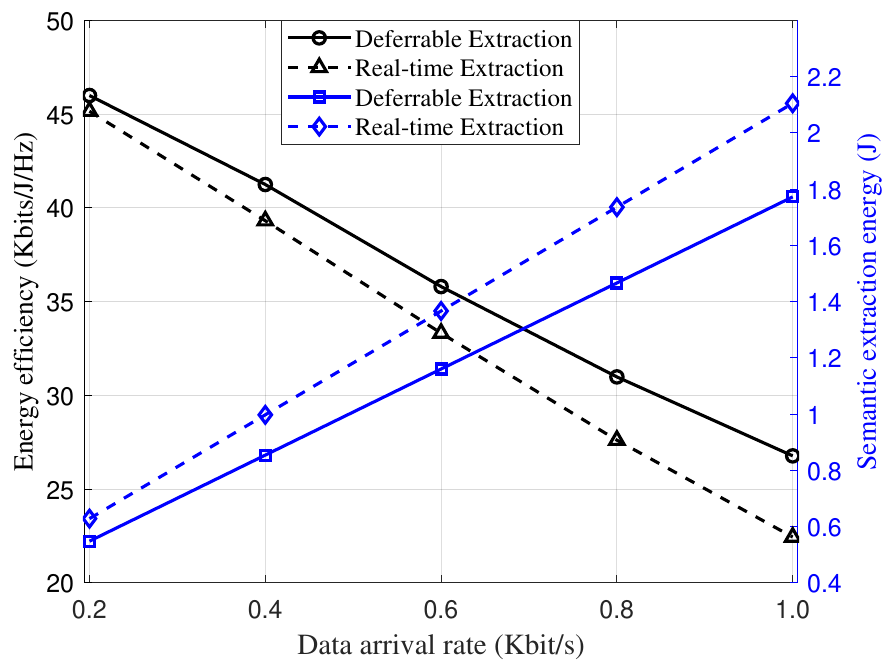}
	\caption{Deferrable extraction improves long-term performance.}\label{performance-deferredcomp}
\vspace{-0.3cm}
\end{figure}

The RIS's different locations also change the channel conditions, influencing both the semantic extraction and transmission processes.  Hence, we evaluate the energy efficiency by varying the RIS's x-coordinate from $1$ to $7$, as shown in Fig.~\ref{fig-ris-settings}(b). We introduce a benchmark scheme that does not include the semantic extraction, denoted by Non-Semantic. {We also compare JTAC with a Lyapunov-based optimization scheme (denoted as Lyapunov). Its main algorithmic principle is to construct a drift-plus-penalty function, which optimizes system performance while maintaining queue stability~\cite{Cang-2024iotj}.} The JTAC scheme demonstrates the best energy efficiency compared to the other schemes. {The Lyapunov-based optimization focuses on per-slot performance improvement and thus achieves lower performance than JTAC in long-term transmission scenarios.} It is also observed that the energy efficiency initially  decreases as the RIS moves away from the AP. However, as it gets closer to the SUs, the energy efficiency begins to increase. This is because a shorter distance between the RIS and either the AP or the SUs contributes to stronger channel gains, which improves  transmission performance and thus provides greater  flexibility to semantic extraction. This observation suggests that the RIS can be deployed closer to the AP or the SUs in practical implementations. An interesting observation is that the Non-Semantic scheme exhibits a larger performance variation as the RIS's location varies. This is because the performance of the Non-Semantic scheme is directly affected by the channel conditions. In contrast, the JTAC scheme can adaptively balance semantic extraction and transmission to cope with varying channel conditions, demonstrating its better robustness in dynamic scenarios.

\begin{figure}[t]
	\centering
	\includegraphics[width = 0.45\textwidth]{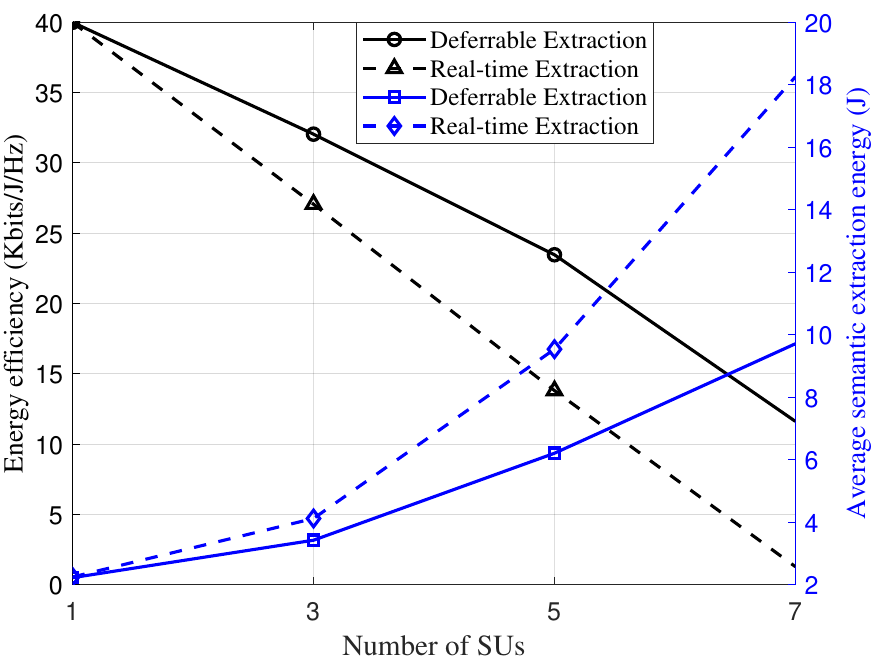}
	\caption{{Energy efficiency under different numbers of SUs.}}\label{num-ee}
\vspace{-0.3cm}
\end{figure}

\begin{figure}[t]
	\centering
	\subfloat[Semantic extraction trend.]{\includegraphics[width=0.22\textwidth]{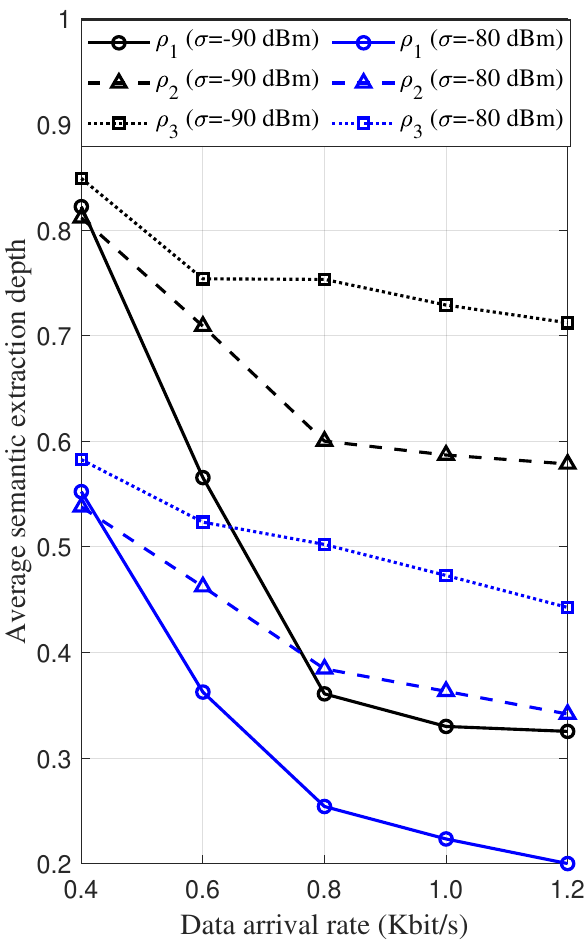}}
	\subfloat[Energy ratio trend.]{\includegraphics[width=0.22\textwidth]{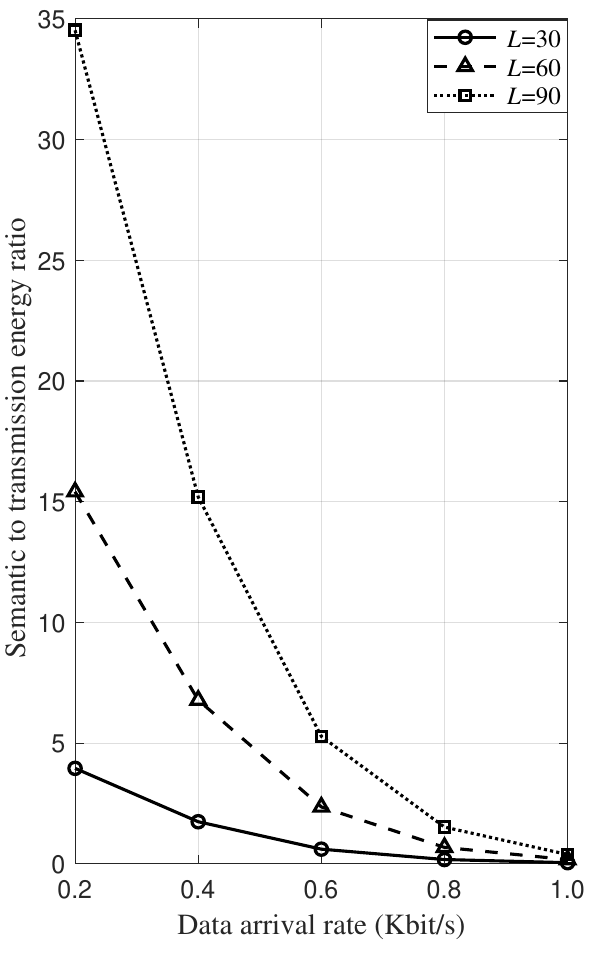}}
	\caption{Semantic decisions with varying traffic demands.}\label{fig-behaviors}
\vspace{-0.2cm}
\end{figure}

\begin{figure*}[t]
	\centering
	\subfloat[Case~\uppercase\expandafter{\romannumeral 1} focuses on semantic extraction.]{\includegraphics[width=0.33\textwidth]{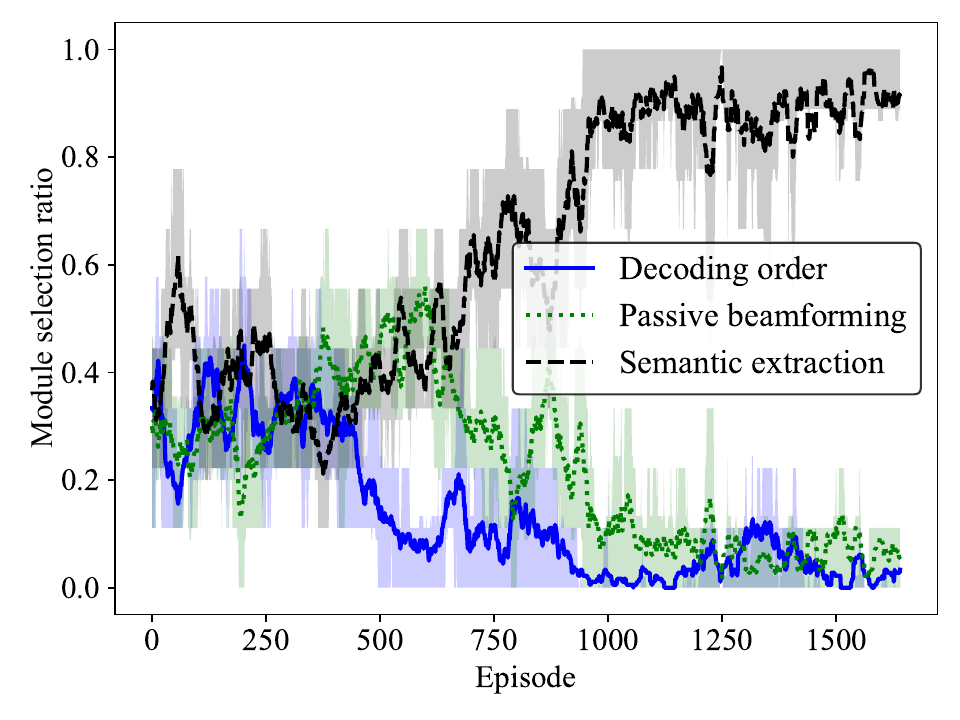}}
	\subfloat[Case~\uppercase\expandafter{\romannumeral 2} focuses on passive beamforming.]{\includegraphics[width=0.33\textwidth]{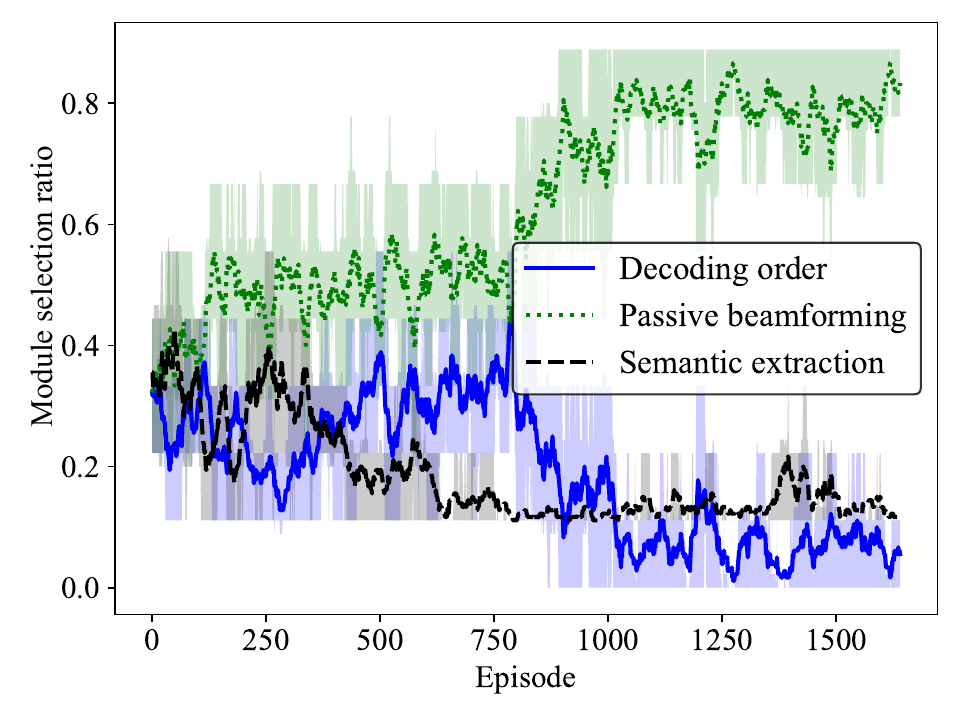}}
	\subfloat[Case~\uppercase\expandafter{\romannumeral 3} focuses on NOMA decoding order.]{\includegraphics[width=0.33\textwidth]{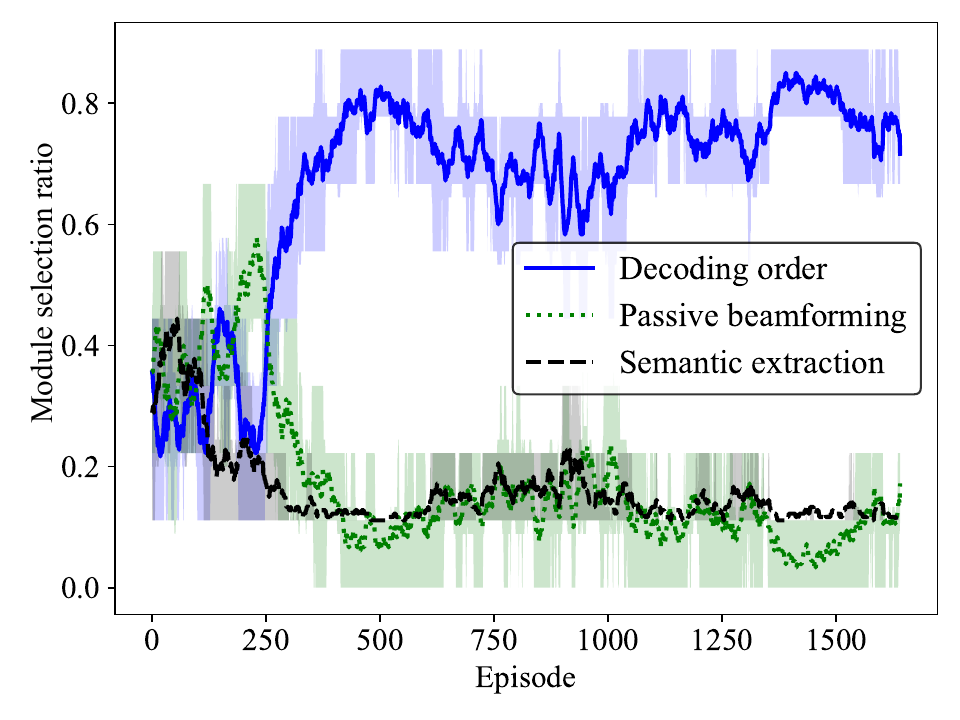}}
	\caption{Adaptive selection of the optimization modules under different cases.}\label{mode-selection-cases}
\vspace{-0.3cm}
\end{figure*}

The deferrable extraction allows the SUs to allocate the extraction tasks over multiple time slots, which aims to balance the traffic demands and  improve the long-term performance. To validate this, we compare the deferrable  semantic extraction scheme with the real-time semantic extraction, as shown in Fig.~\ref{performance-deferredcomp}. When the data arrival rate is low, the energy efficiency improvement of the deferrable scheme is not significant, with only $1.85\%$. The reason is that the SUs have sufficient capabilities to process the newly arrived data in time when they have light traffic demands. However, as the SUs' traffic demands increase, the advantages of the deferrable extraction scheme gradually become more significant, leading to a $19.28\%$ improvement. This is because balancing a heavy extraction task across multiple time slots allows the SUs to better utilize the potentially idle time slots. This is further reflected in the energy consumption of semantic extraction. The deferrable scheme becomes more energy-saving by better utilizing the computational resources over the long term.

We evaluate the scalability of the proposed deferrable semantic extraction scheme with different numbers of SUs, as shown in Fig.~\ref{num-ee}.
The energy efficiency decreases due to intensified access competition, which requires higher power disparities for NOMA transmissions. As such, SUs should perform deeper semantic extraction, and thus lead to a higher average semantic extraction energy consumption. However, the deferrable extraction scheme distributes semantic extraction tasks over multiple slots and allows flexible transmission scheduling, significantly reducing semantic extraction energy consumption and achieving better performance than the real-time extraction scheme.

\subsection{Key Insights and Lightweight Design}
\begin{figure}[t]
	\centering
	\includegraphics[width = 0.45\textwidth]{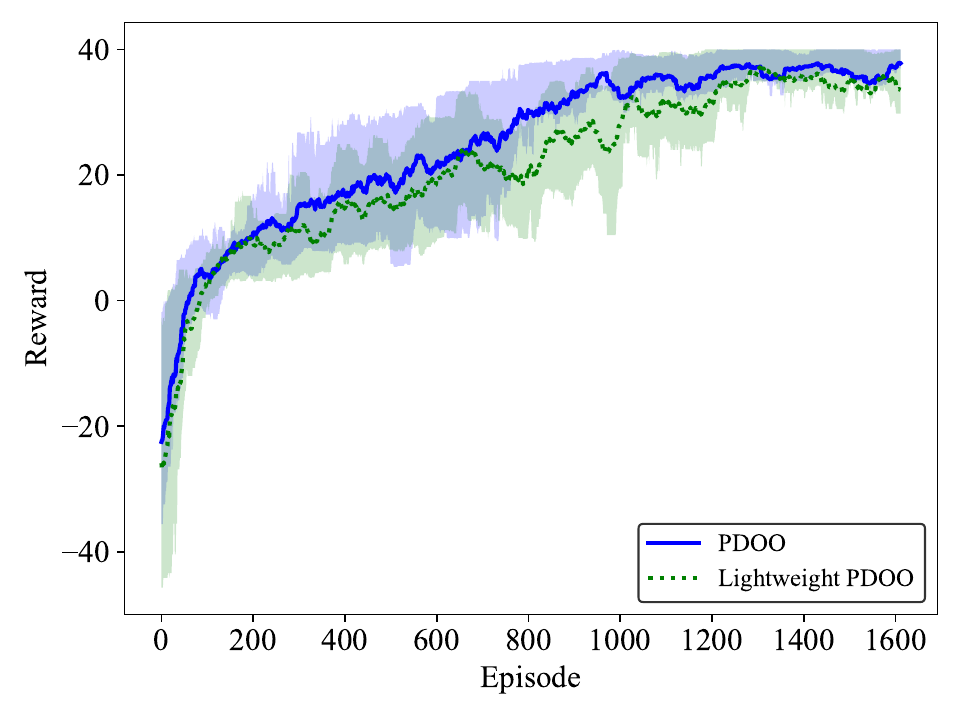}
	\caption{Lightweight designs improve  the learning efficiency.}\label{fig-drl-simplify}
\vspace{-0.4cm}
\end{figure}
Beside the  energy efficiency performance, we also examine how the SUs adjust their semantic extraction under different network parameters. In Fig.~\ref{fig-behaviors}(a), we record the average semantic extraction depth of three SUs over an entire working period. As the SUs' traffic demands increase, all SUs tend to enhance their extraction depth to generate lighter data loads for NOMA transmissions. Meanwhile, it is observed that the differences in the SUs' semantic extraction depth gradually become more significant. This allows the SUs to transmit with diverse power levels, which effectively enhances their SIC decoding and thus enables them to handle more traffic demands. {Furthermore, as the background noise power increases (from $-90$ dBm to $-80$ dBm), it is observed that the SUs adopt deeper semantic extraction. This is because higher noise levels can increase the transmission cost, prompting users to reduce their traffic demand by using a larger semantic extraction depth.}
We further explore the energy trade-off between the semantic extractions and transmissions for the SUs. As shown in Fig.~\ref{fig-behaviors}(b), we present  energy consumption ratio between semantic extraction and transmission as the SUs' traffic demands increase. When the SUs' traffic demands are low, most energy is consumed by semantic extraction. This is because the semantic extraction effectively helps the SUs reduce their transmission loads. However, as the traffic demands gradually increase, even with deeper semantic extraction, the transmission energy consumption increases significantly.  This encourages us to employ a larger RIS to assist the NOMA transmissions in high-demand traffic scenarios.

The optimization selection in DRL module is to identify the most suitable control variable for the optimization module to solve based on the current network status. To validate this, we present the selection proportions of the three optimization methods under three specific network cases as follows. Case~\uppercase\expandafter{\romannumeral 1}: The channel remains stable with only minor traffic fluctuations. Case~\uppercase\expandafter{\romannumeral 2}: The traffic demands remain stable but the channel fluctuates. Case~\uppercase\expandafter{\romannumeral 3}: Both the channel and the traffic experience  high fluctuations. As shown in Fig.~\ref{mode-selection-cases}(a), the RIS's passive beamforming optimization is mostly selected in Case~\uppercase\expandafter{\romannumeral 1}. This is because the SUs do not need to frequently adjust data size or change the decoding order to enhance their transmission capabilities. Instead, we only need to optimize the RIS's passive beamforming to adapt to the real-time channel conditions. Conversely, in Case~\uppercase\expandafter{\romannumeral 2}, when the channel remains stable and only traffic fluctuations occur, the semantic extraction optimization is sufficient for the SUs to adjust their traffic demands to  appropriate levels, as shown in Fig.~\ref{mode-selection-cases}(b). In Case~\uppercase\expandafter{\romannumeral 3}, when both the channel and traffic exhibit high fluctuations, it becomes less effective to adjust either the semantic extraction or the RIS's passive beamforming. Instead, we can dynamically reassign the SUs' transmission priorities by changing their decoding order to better adapt to the current states, as shown in Fig.~\ref{mode-selection-cases}(c).
It is observed that the PDOO efficiently learns the stable optimization selection strategies in different cases. These results verify the capability of the PDOO to select the suitable control variable in the optimization module based on varying network status.
\begin{table}[t]
\centering
\caption{Average running time per step (in seconds).}
\label{tab:runtime}
\begin{tabular}{c|cccc}
\hline
Sizes of RIS $L$ & \textbf{$L=10$} & \textbf{$L=30$} & \textbf{$L=50$} & \textbf{$L=70$} \\
\hline
ALL-Selection      & 0.1625 & 0.2817 & 0.6107 & 1.0273 \\
PDOO               & 0.1069 & 0.1182 & 0.1338 & 0.1857 \\
Lightweight PDOO   & 0.0191 & 0.0195 & 0.0206 & 0.0265 \\
\hline
\end{tabular}
\end{table}

To further reduce the computational complexity and improve the learning efficiency of the PDOO framework, we develop simplified designs for the optimization module (denoted as the Lightweight PDOO), as discussed in Section~\ref{lightweight-pgoo}. The Lightweight PDOO is more suitable for long-term operation in energy-constrained scenarios, such as edge computing systems and wireless sensor networks. Instead of obtaining the precise solutions by solving high-complexity optimization problems, we employ heuristic methods to significantly reduce the computational complexity in each learning round. In Fig.~\ref{fig-drl-simplify}, we compare the reward performance of the PDOO and Lightweight PDOO. Interestingly, the Lightweight PDOO achieves reward performance comparable to that of the PDOO after converging to stable values. However, its running time is significantly reduced to $0.0265$ s/step compared to $0.1857$ s/step for the PDOO. This is due to the powerful learning capability of the PPO, which can effectively compensate for the performance loss caused by the proposed heuristic methods. These observations highlight the potential to design more efficient optimization modules using approximations when developing the PDOO framework.

 {As shown in Table~\ref{tab:runtime}, we also compare the running time of three algorithms, i.e., ALL-Selection, PDOO, and Lightweight PDOO, evaluated on a CPU with the model AMD Ryzen 7 5800H. It is observed that the optimization selection in PDOO significantly reduces the running time compared to the ALL-Selection method, especially when the RIS size $L$ becomes large. Moreover, by designing lightweight methods for the NOMA decoding order and the passive beamforming, the Lightweight PDOO further decreases the running time. Interestingly, when $L$ is small, the variations in running time of PDOO and Lightweight PDOO are less noticeable. This is because optimizing passive beamforming brings limited performance improvement, leading to fewer selections and thus a smaller impact on the overall running time.}

\section{Conclusions}\label{conclusions}
We have investigated an RIS-assisted semantic-aware NOMA wireless network, where the RIS reconfigures the channel conditions while the semantic extraction reshapes the SUs' traffic demands to adapt to scenarios with dynamic network status. To improve long-term transmission performance, we have explored a deferrable semantic extraction scheme. It allows SUs to allocate semantic extraction tasks across multiple time slots to better exploit dynamic wireless resources. To enable a lightweight online optimization, we have proposed the PDOO algorithm. It shows that dynamically optimizing one control variable at a time rather than all variables can effectively reduce computational complexity while maintaining considerable performance in long-term optimization. The numerical results have demonstrated that the deferrable semantic extraction scheme can more efficiently utilize multi-slot resources, thus significantly improving long-term energy efficiency. Furthermore, the PDOO framework effectively improves the learning performance and reduces the running time compared to the benchmark methods. This work adopts a simple linear semantic extraction model to facilitate theoretical analysis. In future work, we will consider more realistic and accurate semantic extraction models tailored to specific application scenarios.

\footnotesize
\bibliographystyle{IEEEtran}
\bibliography{reference}
\end{document}